\documentclass[12pt]{iopart}

\usepackage{graphicx}
\usepackage{cite}
\usepackage[dvipsnames]{xcolor}
\usepackage[margin=1.5cm]{geometry}
\usepackage[normalem]{ulem}

\expandafter\let\csname equation*\endcsname\relax
\expandafter\let\csname endequation*\endcsname\relax
\usepackage{amsfonts}
\usepackage{amsmath}
\usepackage{subfigure}
\def\vep{\varepsilon}
\def\uI{\overline{\overline{I}}}
\def\u0{\overline{\overline{0}}}
\def\uz{\overline{\overline{\zeta}}}
\def\uza{\overline{\overline{\zeta}}\,^{\alpha}}
\def\uze{\overline{\overline{\zeta}}\,^{e}}
\def\uzep{\overline{\overline{\zeta}}\,^{\prime\, e}}
\def\uzepp{\overline{\overline{\zeta}}\,^{\prime\prime\,e}}

\def\uzpp{\overline{\overline{\zeta}}\,^{\prime\prime}}
\def\ua{\overline{\overline{\alpha}}}
\def\um{\overline{\overline{\mu}}}

\def\umpp{\overline{\overline{\mu}}\,^{\prime\prime}}
\def\uep{\overline{\overline{\vep}}}

\def\ueppp{\overline{\overline{\vep}}\,^{\prime\prime}}

\def\ubGt{\overline{\overline{\textbf{\c{G}}}}}
\def\ubO{\overline{\overline{\bf{O}}}}
\def\uiO{\overline{\overline{\textbf{$\iota$}}}}
\def\ubG{\overline{\overline{\bf{G}}}}
\def\ucG{\overline{\overline{\c{G}}}}

\def\ubA{\overline{\overline{\bf{A}}}}
\def\ubB{\overline{\overline{\bf{B}}}}
\def\ubC{\overline{\overline{\bf{C}}}}
\def\ubD{\overline{\overline{\bf{D}}}}

\def\uA{\overline{\overline{A}}}
\def\uB{\overline{\overline{B}}}
\def\uC{\overline{\overline{C}}}
\def\uD{\overline{\overline{D}}}
\def\uG{\overline{\overline{G}}}
\def\uH{\overline{\overline{H}}}
\def\uAa{\overline{\overline{\mathfrak{A}}}}
\def\uP{\overline{\overline{\mathfrak{P}}}}

\def\ue{\overline{\overline{e}}}
\def\p{\partial}
\def\pr{\prime}

\def\Dom{\triangle\omega}
\def\vq{\vec{q}}
\def\vS{\vec{S}}
\def\vF{\vec{F}}

\def\vX{\vec{\Xi}}
\def\vr{\vec{r}}
\def\vR{\vec{R}}
\def\vro{\vec{\rho}}
\def\vvro{\vec{\varrho}}
\def\vk{\vec{k}}
\def\vJ{\vec{J}}

\def\vbF{\vec{\bf{F}}}
\def\vbJ{\vec{\bf{J}}}
\def\vbT{\vec{\bf{T}}}
\def\vbU{\vec{\bf{U}}}
\def\zt{\zeta_{t}}
\def\zz{\zeta_{z}}
\def\ept{\varepsilon_{t}}
\def\epz{\varepsilon_{z}}
\def\mt{\mu_{t}}
\def\mz{\mu_{z}}

\def\DA2{\triangle^{2}_{A}}
\def\kt2{k_{t}^{2}}
\def\kz2{k_{z}^{2}}
\def\htt{\hat{t}}
\def\hs{\hat{s}}
\def\hz{\hat{z}}

\def\uu{\overline{\overline{u}}}

\def\Re{\mathfrak{R}}
\def\Im{\mathfrak{I}}
\def\Tr{{\rm Tr}} 

\begin{document}

\title[Radiative-Conductive Heat Transfer Dynamics]{Radiative-Conductive Heat Transfer Dynamics in Dissipative Dispersive Anisotropic Media}

\author{Hodjat Mariji$^1$\footnote{~Corresponding author.} \& Stanislav Maslovski$^2$}

\address{$^1$Independent Researcher, {\rm previously affiliated with} Instituto de Telecomunica\c{c}\~{o}es, Universidade de
	Coimbra (p\'{o}lo II), 3030-290 Coimbra, Portugal}
\ead{astrohodjat@gmail.com}

\address{$^2$Instituto de Telecomunica\c{c}\~{o}es and Dept. of Electronics, Telecommunications and Informatics, University of Aveiro, Campus Universit\'{a}rio de Santiago, 3810-193 Aveiro, Portugal}
\ead{stanislav.maslovski@ua.pt}

\vspace{10pt}
\begin{indented}
\item[]October 2024
\end{indented}

\begin{abstract}
We develop a self-consistent theoretical formalism to model the dynamics of
heat transfer in dissipative, dispersive, anisotropic nanoscale media, such as
metamaterials. We employ our envelope dyadic Green’s function method to
solve Maxwell’s macroscopic equations for the propagation of fluctuating
electromagnetic fields in these media. We assume that the photonic radiative
heat transfer mechanism in these media is complemented by dynamic phononic
mechanisms of heat storage and conduction, accounting for effects of local heat
generation. By employing the Poynting theorem and the fluctuation-dissipation
theorem, we derive novel closed-form expressions for the radiative heat flux
and the coupling term of photonic and phononic subsystems, which contains
the heating rate and the radiative heat power contributions. We apply our
formalism to the paraxial heat transfer in uniaxial media and present relevant
closed-form expressions. By considering a Gaussian transverse temperature
profile, we also obtain and solve a system of integro-differential heat diffusion
equations to model the paraxial heat transfer in uniaxial reciprocal media. By
applying the developed analytical model to radiative-conductive heat tranfer in
nanolayered media constructed by layers of silica and germanium, we compute
the temperature profiles for the three first orders of expansion and the total temperature
profile as well. The results of this research can be of interest in areas of science and technology related to thermophotovoltaics, energy harvesting,
radiative cooling, and thermal management at micro- and nanoscale.
\end{abstract}


\section{Introduction}

It is well-known that the heat transfer (HT) in media \footnote{~In  this work, we do not consider fluids for which the convection heat transfer should be included.} is mediated by
two main mechanisms: photonic, the radiative heat transfer (RHT), and
phononic, the thermal conduction. The RHT is associated with the
fluctuating electromagnetic fields (FEFs) which are produced by
thermally agitated random motions of electric charges in the medium
while the thermal conduction is associated with collective
oscillations of atoms in a lattice. The phononic mechanism comprises
the thermal heat conduction and storage, which has been known for a
long time (e.g., see \cite{JAMS-1}). The RHT is governed by the
fluctuation-dissipation theorem (FDT) and Rytov's fluctuational
electrodynamics \cite{1}. It has been also theoretically and
experimentally shown \cite{2, 3, 4, 5, 6, 7, 8, 9, 10, PRB-2023, Nature-2024} that the
radiative heat exchange becomes dominating when separation between
objects is decreased. While the far-field RHT for two separated
objects is limited by the Planck's blackbody radiation law, the RHT in
the near field can increase by orders of magnitude above the blackbody
radiation limit when the separation becomes deeply sub-wavelength. For
such case, it is necessary to consider the near-filed radiative heat
transfer (NFRHT) \cite{11} since near-field effects, i.e. the
collective influence of diffraction, interference, and tunneling of
waves, become important \cite{12}. Hence, the Planck's blackbody
radiation law and the Stefan-Boltzmann law are no longer valid to
determine the spectral emissivity and total emissive power when
object's geometric dimensions are comparable to the characteristic
wavelength of thermal radiation.

Nowadays, there is a growing interest in the study of NFRHT in the
micro- and nano-scale objects, especially, metamaterials (MMs). Due to
their peculiar properties, MMs are increasingly adopted in many
scientific and technological applications \cite{JAMS-2, 13, 14, 15,
	16, 17, 18, 19, 20, 21, 22, 23, 24, 25, 26, 27, 28, 29, 30,
	smartNFRHT, MMNat-2024, MMApp-2023}.  Recently, the super Planckian NFRHT has attracted
attention of science and technology \cite{31, 32, 33, 34, 35, 36, 29,
	38, SupP-2022, SupP-2024}
	with applications in thermionic systems, thermophotovoltaics,
energy harvesting, local thermal control and management, thermal
modulation, nanoscale infrared imaging and mapping, and
nanomanufacturing. This development has been complemented by studies
of thermal properties of MMs \cite{39, 21, 41, 30, 43, 44, 45, 46, 47,
	48, SR-2022, MP-2024}. 
	Moreover, further extent of nanoscale RHT studies into
multidisciplinary areas may assist in solving current societal
challenge. For instance, regarding the increasing dangers of global
warming, new MMs have been proposed for applications in
thermophotovoltaics systems that can efficiently convert thermal
energy from any kind of heat source into electricity \cite{49} or
perform radiative cooling under direct sunlight, aiming to reduce
global energy consumption \cite{50}. All-dielectric MMs for similar
energy harvesting applications have been designed by using Machine
Learning techniques and employing artificial intelligence for mapping 
information and light in metaphotonics \cite{51, AI-MP}. It is worth to mention that control of
thermal flux by utilizing photonic devices becomes increasingly
attractive because of applications of such devices as thermal diodes
and thermal logic gates, and even as thermal cloaks \cite{43, 52, 53, 54, 55}. Another actively developing area is the microfluidic and
nanofluidic thermal management for electronic devices \cite{MicroTM, MicroTM1-2024, MicroTM2-2024, NanoTM}. All these observations highlight the necessity of paying
attention to the mechanisms of combined conductive-radiative HT in
nanoscale media such as MMs.

Although many important theoretical methods have been proposed to
analyze the heat diffusion in MMs based on the fluctuational
electrodynamics and the microscopic/ macroscopic Maxwell equations,
\cite{2, 3, 4, 5, 6, 7, 24, 29, 56, 57, 58, 59, Ito-2017}, there is not much
theoretical work specifically devoted to modeling the
\textit{dynamics} of HT (DHT) in nanoscale media. Recently, in the
framework of the slowly varying amplitude (SVA) approach, we presented
a novel method, namely, the envelope dyadic Green's function (EDGF)
\cite{60}, which is different from the standard dyadic Green’s
function method. With this method, one can obtain the propagating
fluctuating electromagnetic fields in dispersive anisotropic media by
using the precalculated EDGFs for the uniaxial MMs.

In this paper, based on our EDGF method \cite{60}, we aim to develop a
self-consistent theoretical formalism to model the DHT in dissipative
dispersive anisotropic nanoscale media (DDANM) such as MMs, with
dyadic constitutive parameters such as the permittivity and the
permeability. The formalism presented in this article neglects
magneto-electric interactions that exist in bianisotropic media,
however, it can be extended to include bianisotropic effects
\cite{SPIE-Stanislav}.

In what follows, assuming that the photonic HT mechanisms are
complemented by dynamic phononic mechanisms of heat storage and
conduction, we shall present closed form relations for the coupling
term of the photonic-phononic HT mechanisms, expressed in terms of the
heating rate and the radiative heat flux power. As we mentioned
above, these relations are obtained using our EDGF method by employing
the Poynting and FDT theorems, while accounting for the effects of
local heating and the material phase transitions. At the next step, we
apply our formalism to paraxial propagation in uniaxial media such as
one-dimensional crystalls of nanolayers with an aim to solve the heat
diffusion equations in such media. To do this, we express the
transverse temperature profile as a superposition of multiplicative
terms comprising Gaussian and polynomial functions and present a group
of integro-differential equations for the temperature distribution in
the axial direction. Then, the formulated equations are solved by two complementary methods.

This article is organized as follows: In Section 1, we explain the
fundamental concepts in the context of heat transfer and present
relevant formulas. In Section 2, by considering the propagation of
fluctuating electromagnetic fields with SVAs (FEFSVAs) in reciprocal DDANM, we
construct a self-consistent DHT formalism and obtain closed-form
expressions for the heating rate, the radiative power, and the
radiative heat flux. In Section 3, we apply our DHT formalism
for the propagation of paraxial FEFSVAs in a DDAN uniaxial medium. In
Section 4, we obtain and solve the integro-differential equations for the heat
diffusion in such a medium. Finally, in Section 5, we discuss
the obtained results and make conclusions.

\section{Heat transfer}
In the modeling of DHT in DDANM, the key unknown is the temperature
distribution that must be extracted from the heat diffusion equation
in the medium which, obviously, combines both phononic and photonic HT
mechanisms. Thermally-agitated noise in the form of FEF is produced by
random motions of charge carriers and vibrations of polarized atoms in
the medium. In turn, FEF when propagating in a lossy medium,
dissipates electromagnetic energy into the heat in the form of similar
random vibrations. Thus, random phononic processes in a material with
loss are coupled to random photonic processes and vice versa.

The phononic mechanism mentioned above comprises the heat conduction
and storage due to the processes such as lattice vibrations, material
phase transitions, etc. In this section, we consider the effective
thermal conduction of the materials and the latent heat associated
with such processes. On the other hand, in regard to the photonic HT,
here we need to calculate the generated heat term which must occur in
the heat diffusion equation.

The equation for the heat diffusion due to phonons in a medium kept
under constant volume can be constructed phenomenologically, by
balancing the growth rate of the phononic thermal energy per unit
volume, $dU_{V}/dt$, with the rates of heat supply, $dQ^{s}_{V}/dt$,
and release, $dQ^{r}_{V}/dt$, per unit volume, as follows:
\begin{equation}\label{1}
	{dU_{V} \over dt} = {dQ^{s}_{V} \over dt} - {dQ^{r}_{V} \over dt}.
\end{equation}
As is well-known from thermodynamics, a variation of the thermal
energy of the medium under isochoric conditions can be expressed
through its temperature variation, $dT$, the specific heat capacity of
the material at constant volume, $c_v(T)$, and the variation of the
latent heat, $dL_p$, associated with material phase transitions.  Here
we consider only the processes in which the change of material mass
density $\varrho_{m}$ is either naturally negligible (such as
ferromagnetic-paramagnetic transitions, etc.) or vanishes due to the
enforced isochoric conditions. Thus, we have
\begin{equation}\label{2}
	dU_{V}= \varrho_{m} c_{v}(T)dT + dL_{p},
\end{equation}
During a phase transition that happens at the transition temperature
$T = T_p$, the material temperature stays constant and thus $dT = 0$,
while $dL_p \neq 0$. When $T \neq T_p$, $dL_p = 0$. At the point
$T = T_p$, the specific heat capacity $c_v(T)$ can be discontinuous,
because $c_v$ can differ for the two phases.

In order to calculate the rate of heat supply that drives the phononic
HT subsystem, we need to consider the heating rate density due to the
passing radiation, $h^\Sigma_{EM}$, where the label $^\Sigma$ means
that contributions from FEFs with all possible frequencies are taken
into account, and the density of the generated heat from other
sources, $g_c$, which can be chemical processes, etc. Thus, the heat
supply rate per volume of the medium reads
\begin{equation}\label{3-hs}
	{dQ_V^s \over dt} = h_{EM}^\Sigma + g_c.
\end{equation}
In order to find the heat release rate, we have to consider the
phononic heat flux density, $\vq_{c}$, and the power density of locally
generated thermal radiation, $p^\Sigma_{rad}$:
\begin{equation}\label{3-rad}
	{dQ_V^r \over dt} = \nabla\cdot\vq_{c} + p^\Sigma_{rad},
\end{equation}
where $\vq_{c} = \ua_{c}\cdot\nabla T$, with $\ua_{c}$ as the phononic
thermal conductivity tensor of the medium.

The heat diffusion equation associated with phononic processes can
be written as follows:
\begin{equation}\label{3-HDE}
	\varrho_{m} c_{v}(T){dT\over dt} + {dL_{p} \over dt} =-\nabla\cdot({\ua_{c}\cdot\nabla T}) + g_c + (h_{EM}^\Sigma - 	p^\Sigma_{rad}),
\end{equation}

In order to quantify the coupling term
$(h_{EM}^\Sigma - p^\Sigma_{rad})$ at every carrier frequency
$\omega$, the heating rate density, $h_{EM}(\omega)$, and the
power density of locally generated thermal radiation, $p_{rad}(\omega)$, must be obtained based on the FDT and the electromagnetic properties of the material. In addition, for computing the total HT rate that is the sum of the conductive HT rate, which is the first term in the right-hand side of Eq.~(\ref{3-HDE}), and the radiative HT rate (not present in Eq.~(\ref{3-HDE})), we need to calculate the radiative heat flux
density associated with the FEF at the same frequency,
$\vq_{rad}(\omega)$. We postpone quantifying these terms untill
Section 2, however, we briefly point out the main physical concepts in the
contexts involving conversion of the radiation energy into mechanical work and
vice versa based on the microscopic and macroscopic views of the
modeling of DHT in DDANM.

In the microscopic view, we need principally to consider the free
energy of the system (involving FEFs) which is a fair thermodynamic
potential comparing to the total energy. Free energy, as a measure of
work that a system can do at constant temperature, takes into account
disorder in the system through including entropy and is expressed as a
difference between the change in total energy and the energy lost in
the form of heat. Given the photonic states density and the photonic
distribution function (Bose-Einstein relation) with the known chemical
potential of radiation (through the relationship with the chemical
potentials of carriers), one can obtain the entropy of radiation in
order to calculate the free energy in both equilibrium and
non-equilibrium cases \cite{Landau}. However, we reserve such a
microscopic view for a future work and, in the following, we proceed
with a macroscopic view by considering only the mean squares of
thermal FEFs and their relation to the total energy and the power
flux density.

In order to model the DHT in DDANM macroscopically, we employ the
Poynting's theorem for FEFs, which gives the work done by these fields
in the medium. In other words, we assume that the heating rate
density, $h_{EM}$, in the presence of FEFs can be determined from the
Poynting's theorem for macroscopic electromagnetic fields and the
material relations for such fields, in the same way as for
non-stochastic electromagnetic processes (for the stochastic heat
diffusion study, see, e.g., \cite{JAMS-3}). For this assumption to
hold, the characteristic spatial scale of phononic thermal
fluctuations generated by the dissipative processes (e.g., electron
and phonon scattering on impurities, etc.) must be much smaller than
that of the electromagnetic fluctuations, which is a reasonable
assumption when considering macroscopic electromagnetic fields.

In order to obtain the coupling term and heat flux in the framework of
the Poynting's theorem, we need to calculate the work done by FEFs,
$\vF^{\,\alpha} = \vF^{\,\alpha}_{0}\cos(\omega t+\phi_{\alpha})$,
where $\phi_{\alpha}$ is an arbitrary phase with
$\alpha = e, m$.\footnote{~Hereafter, the superscripts/subscripts of
	$e$ and $m$ in a quantity, in respective, denotes the electric and
	magnetic components of the quantity.} Since in what follows we
intend to work with the oscillating fields whose amplitudes and phases
fluctuate, it is more convenient to write the electric and magnetic
fields as
$\vF^{\,\alpha} = \Re \left( \vF^{\,\alpha}_{m} e^{-i\omega t}\right)$, where $\vF^{\,\alpha}_{m}$ are the time-dependent complex phasors
of the fluctuating electric and magnetic fields and $\Re(\cdot)$
denotes the real part of a complex quantity. Moreover, $\vF^{\,e}$ and
$\vF^{\,m}$, the fluctuating electric and magnetic fields,
respectively, are related to the fluctuating electric and magnetic
induction vectors, $\vX^{\,e}$ and $\vX^{\,m}$, respectively, through
the constitutive relations of the form
\begin{equation}\label{4}
	\vX^{\,\alpha}=\uza\cdot\vF^{\,\alpha}, \qquad \alpha = e, m
\end{equation}
where the dyadics $\uz\,^e$ and $\uz\,^m$ are, respectively, the
dyadics of absolute permittivity, $\uep$, and absolute permeability, $\um$, of
the medium. When dealing with time-dependent fields in dispersive
media, $\uza$ can be considered as a dyadic integro-differential
operator. It should be noted that the complex field phasors,
$\vF^{\,e}_{m}$, $\vF^{\,m}_{m}$, $\vX^{\,e}_{m}$, and
$\vX^{\,m}_{m}$, also satisfy the frequency-domain Maxwell equations
of the following form
\begin{equation}\label{8}
	\nabla\cdot\vX^{\,e[m]}_{m}=\rho_{m}^{e[m]}, \qquad \nabla\times\vF^{\,e[m]}_{m} =-[+]\left( (\p_{t} - i\omega)\vX^{\,m[e]}_{m} + \vJ_{m}^{\,m[e]}\right),
\end{equation}
where $\vJ_{m}^{\,e[m]}$ and $\rho_{m}^{\,e[m]}$ are, respectively, the complex amplitudes of the electric
(magnetic) current density, $\vJ^{\,e[m]}$ and the electric (magnetic) charge density, $\rho^{\,e[m]}$.\footnote{~We use shorthand
	notation $\alpha[\beta]$ for respective selection of either $\alpha$
	or $\beta$, where $\alpha$ and $\beta$ are isolated terms or groups
	of indices. For example,
	$\nabla\cdot\vX^{\,e[m]}_{m}=\rho_{m}^{e[m]}$ means
	$\nabla\cdot\vX^{\,e}_{m}=\rho_{m}^{e}$ and
	$\nabla\cdot\vX^{\,m}_{m}=\rho_{m}^{m}$.}  By keeping the time
derivative of the complex field amplitudes, $\p_{t} \Xi_{m}$, in the
above equations, we emphasize that these quantities may vary with time
as we expect in the fluctuating fields case.

The work $W_{EM}$ of embedded electric and magnetic current sources
on exciting such random time-harmonic fields (per a period of
oscillations) reads
\begin{align}\label{5-work}
	{\omega W_{EM}\over 2\pi}&= \left\langle\int P_{src} dV\,\right\rangle_T \nonumber \\
	& = \left\langle\int
	\left[\nabla\cdot\vS+{1 \over
		2}(\vF^{\,e}\cdot\p_{t}\vX^{\,e}
	+\vF^{\,m}\cdot\p_{t}\vX^{\,m} )
	\right] dV\right\rangle_T,
\end{align}
where $\langle\ldots\rangle_T$ denotes averaging over a period of
oscillations, $\p_{t} = {\p \over \p t}$, and $P_{src}$, the source
power density, is defined by
\begin{equation}\label{5-src}
	\left\langle P_{src}\right\rangle_T=-\left\langle\vF^{\,e}\cdot\vJ^{\,e\,*} + \vF^{\,m\,*}\cdot\vJ^{\,m}\right\rangle_T = -{1 \over 2} \Re\! \left[\vF^{\,e}_{m}\cdot\vJ_{m}^{\,e\,*} + \vF^{\,m\,*}_{m}\cdot \vJ_{m}^{\,m}\right],
\end{equation} 
On the right-hand side of Eq.~(\ref{5-work}), the Poynting
vector $\vS$ that appears in the first term is given by
\begin{equation}\label{5-poynt}
	\left\langle\vS\,\right\rangle_T=\left\langle\vF^{\,e}\times\vF^{\,m}\right\rangle_T = {1 \over 2} \Re\!\left(\vF^{\,e}_{m}\times\vF^{\,m\,*}_{m} \right).
\end{equation}
and the second term of the same expression is the total rate of the
electromagnetic energy dissipation and storage per unit volume of the
medium, which can be written in terms of the phasors as
\begin{equation}\label{7}
	q_{_{EM}}={1 \over 2} \Re\! \left[\vF^{\,e}_{m}\cdot(\p_{t} + i\omega)\vX^{\,e\,*}_{m} + \vF^{\,m\,*}_{m}\cdot(\p_{t}- i\omega)\vX^{\,m}_{m} \right].
\end{equation}

In Eqs.~(\ref{5-work}-\ref{7}), we need to know the FEFs in the
DDANM with known material parameters. In this work, we use our EDGF
method \cite{60} to solve the Maxwell macroscopic equations,
Eq. (\ref{8}).

\noindent In the EDGF formalism, we consider two main assumptions:
\begin{enumerate}
	\item The time-harmonic FEFs $\vF^{\alpha}_{m}$, flux densities
	$\vX^{\alpha}_{m}$, and currents $\vJ^{\alpha}_{m}$ are complex SVAs \cite{60}: 
	\begin{equation}\label{f_sva}
		\begin{aligned}
			\vec{f}(t) &= {1 \over 2}\vec{f}_{m}(t) e^{-i\omega_{0}t} + c.c., \\
			\vec{f}_{m}(t)&={1 \over \pi} \int\limits_{-\Delta\omega/2}^{\Delta\omega/2}\vec{f}(\omega_{0}+\Omega)e^{-i\Omega t}d\Omega, \qquad  f = F, \Xi, J,
		\end{aligned}
	\end{equation}
	where $c.c.$ stands for the complex conjugate of the previous term. Here,
	$\Omega=\omega-\omega_{0}$ with $\omega_{0}$ as the carrier frequency
	and $\Delta\omega \ll \omega_{0}$.
	
	\item The frequency-dependent material dyadics $\uza(\omega)$ [the
	permittivity $\uep(\omega)$ or the permeability $\um(\omega)$]
	behave smoothly near $\omega_{0}$, so that we can expand $\uza$
	around $\omega_{0}$ as follows
	\begin{equation}\label{11}
		\uza(\omega)=\uza_{_{\omega_0}} + (\p_{\omega}
		\uza)_{_{\omega_0}}\Omega + o(\Omega),
	\end{equation}
	where the subscript $\omega_0$ emphasizes that the quantity has to be calculated at $\omega_{0}$ (we keep this notation in the rest of this paper).
\end{enumerate}
Based on these assumptions, we obtain a relation (which will be used
in the next section) between the complex SVAs of the fluctuating
electromagnetic fields and flux densities. Regarding Eqs.~(\ref{4}) and
(\ref{f_sva}), we have
\begin{equation}\label{15-extra}
	\vX^{\,\alpha}_{m}(t)= {1 \over \pi}\!\!\int\limits_{-\Delta\omega/2}^{\Delta\omega/2}\!\!\uza(\omega_0+\Omega)\cdot \vec{F}^{\,\alpha}_m(\omega_0+\Omega)e^{-i\Omega t} d\Omega.
\end{equation}
Now, using Eqs.~(\ref{f_sva}-\ref{11}) and keeping only the
first-order term of the Taylor's expansion over $\Omega$, we obtain
\begin{equation}\label{15-extra-extra}
	\vX^{\,\alpha}_{m}(t)= \uza\cdot\vec{F}^{\,\alpha}_{m} + (\p_{\omega}\uza)_{_{\omega_0}}\!\!\cdot\left[{1 \over \pi}\!\!\int\limits_{-\Delta\omega/2}^{\Delta\omega/2}\!\!\!\Omega\,\vec{F}^{\,\alpha}_{m}(\omega_{0}+\Omega)e^{-i\Omega t}d\Omega\right],
\end{equation}
where the bracket in the second term of the right-hand side of
Eq.~(\ref{15-extra-extra}) is the time derivative of
$\vec{F}^{\,\alpha}_{m}(t)$ divided by a factor of $-i$. Thus, we find
the following useful relation between the complex SVAs of the
fluctuating electromagnetic fields and flux densities
\begin{equation}\label{15}
	\vX^{\,\alpha}_{m}(t)=\left[\uza + (\p_{\omega}\uza)(i\p_{t})\right]_{_{\omega_0}}\!\!\cdot\vec{F}^{\,\alpha}_{m},	\qquad \alpha = e\ \mbox{or}\ m.
\end{equation}

Let us introduce the 6-vector electromagnetic fields and currents as follows:
\begin{equation}\label{6_vec}
	\vbF_{m} = 
	\begin{bmatrix}
		\vF^{\,e}_{m}\\
		\vF^{\,m}_{m}
	\end{bmatrix}, \qquad
	\vbJ_{m} = 
	\begin{bmatrix}
		\vJ^{\,e}_{m}\\
		\vJ^{\,m}_{m}
	\end{bmatrix}.
\end{equation}

	$\vbF_{m}$ is given by the following operator form \cite{60}:
	\begin{equation}\label{17}
		\vbF_{m} = \hat{\ubG}\cdot\vbJ_{m},
	\end{equation}
	which is understood as the following integral
	\begin{equation}\label{18}
		\vbF_{m}\left(\vr,t\right)=\int\ubG(\vr,t; \vr^{\,\prime},t^{\prime})\cdot\vbJ_{m}(\vr^{\,\prime},t^{\prime})d^{3}r^{\,\prime}dt^{\prime},
	\end{equation}
	where $\ubG$ is EDGF.

To model DHT, in addition to the extracted FEFs from the EDGF and the
Poynting theorem, we use FDT which states that the fluctuation
properties of a system in thermal equilibrium can be expressed in
terms of the linear response of the same system to an external
perturbation \cite{1}.

According to the FDT, for an anisotropic local medium at local thermal equilibrium, the correlation function of thermally generated random currents reads 
\begin{equation}\label{FDT-0}
	\left\langle\vJ_\omega^{\,\alpha}(\vr,\omega)\vJ_\omega^{\,\alpha}\,^*(\vr^{\,\pr}, \omega^{\,\pr})\right\rangle = 
	{4\Theta (\omega,T) \over \pi} \Im(\uza_\omega)\,\delta(\omega-\omega^{\pr})\delta(\vr-\vr^{\,\prime}),
\end{equation}
where $\vJ_\omega^{\,\alpha}$ is the time-harmonic fluctuating
(electric or magnetic) current density, $\Theta (\omega,T)= \hbar\omega(e^{\hbar\omega/k_{B}T}-1)^{-1}$ is the mean Planck's oscillator energy as a function of frequency and temperature and $\Im (\uza_\omega)$ the imaginary part of the Fourier transform of $\uza$.

\begin{figure}[hbt!]
	\centering
	\includegraphics[scale=.57]{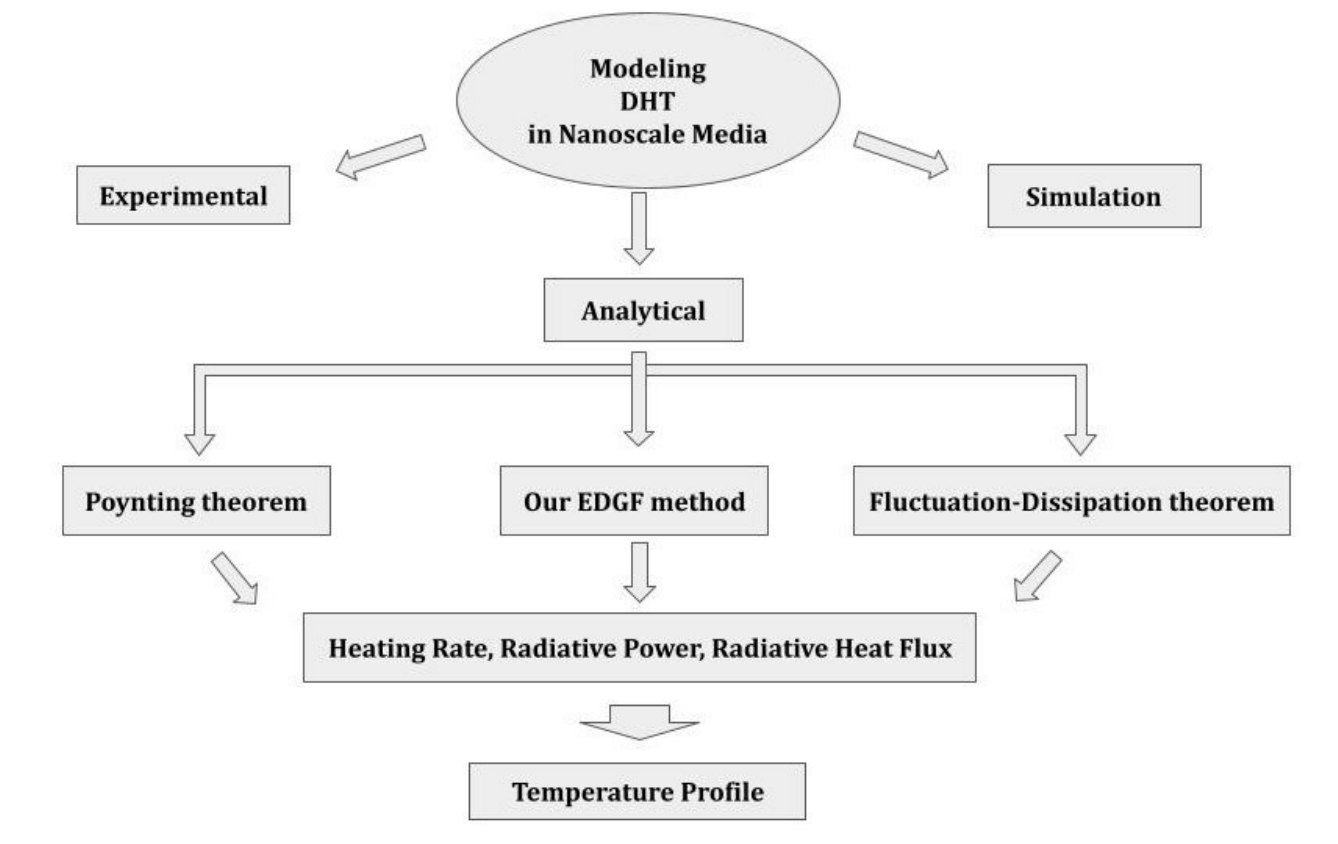}
	\caption{A chart illustrating the position of current work
		among other known DHT modeling approaches.}
	\label{1stfig}
\end{figure}

We conclude this section with a graphical illustration that clarifies
positioning of our method of characterizing the heat transfer dynamics in
DDANM with respect to other known
methods. According to the literature, there are three main approaches,
namely, {\it experimental}, {\it analytical}, and {\it numerical simulation}, to
do this. As presented by the central path of the flowchart shown
in Fig.~ (\ref{1stfig}), the {\it analytical} model of DHT in such media
developed in our work is founded on the three pillars: {\it The Poynting
	theorem, the FDT, and the EDGF method,} which allow us to simultaneously
characterize the {\it heating rate, radiative power, and radiative heat
	flux,} and, with that information, obtain the resulting temperature
distribution within the material, as will be shown in the following
sections.

\section{Heating rate, radiative power, and radiative heat flux}
In this section, in the framework of the EDGF method \cite{60} 
and employing the Poynting theorem and FDT, we obtain closed-form expressions for $h_{EM}$, $p_{rad}$, and $\vq_{rad}$, at
an arbitrary point in the radiation frequency spectrum, $\omega_{0}$,
which we call {\it the carrier frequency}. Next, by integrating over
all carrier frequencies we obtain the total heating rate
$h_{EM}^{\,\Sigma}$, the total radiative power $p_{rad}^{\,\Sigma}$, and the total radiative heat flux $\vq_{rad}^{\,\Sigma}$. It should be noted that we shall calculate the Poynting vector of FEFSVA as a function of carrier frequency ($\omega_{0}$), position, and time for an arbitrary plane inside a medium by obtaining the component of the Poynting vector perpendicular to this plane. Thus, with the yielded quantity it will be possible to describe the RHT in the DDANM with non-uniform distribution of temperature.  It is worth to mention that when employing FDT, which allows us to relate the ensemble average of the square of fluctuating electric (magnetic) current density to the Planck's oscillator energy and the imaginary part of the permittivity (permeability) tensor, we assume that there is no correlation between the electric and magnetic components of the fluctuating currents.

In order to calculate the heating rate density, the coupling term,
$h_{EM}^\Sigma - p_{rad}^\Sigma$, and the heat flux,
$\vq_{rad}^{\,\Sigma}$, by use of the FDT, let us have a closer look at
Eq.~(\ref{7}):
\begin{equation}\nonumber
	q_{_{EM}}={1 \over 2} \Re\! \left[\vF^{\,e}_{m}\cdot(\p_{t} + i\omega)\vX^{\,e\,*}_{m} + \vF^{\,m\,*}_{m}\cdot(\p_{t}- i\omega)\vX^{\,m}_{m} \right].
\end{equation}	
We claim that the right-hand side of this equation is a sum of the
dissipation power (i.e. the rate of transformation of the
electromagnetic energy into heat) and the rate of electromagnetic
energy storage in a unit volume of the medium. Separation of these
contributions into two explicit terms is possible only in very special
cases. On the other hand, for truly monochromatic processes, there is
no energy storage on average, and thus this term is entirely
dissipative.  Likewise, for the processes with finite duration in
which the field amplitude first gradually increases from zero to some
value and then decreases back to zero, the integral
\begin{align}\label{106}
	Q_{EM} &= {1 \over 2} \Re\! \int\limits_{-\infty}^{\infty}\left[\vF^{\,e}_{m}\cdot(\p_{t} + i\omega)\vX^{\,e\,*}_{m} + \vF^{\,m\,*}_{m}\cdot(\p_{t}- i\omega)\vX^{\,m}_{m}\right]_{_{\omega_0}}\!\!\!dt \nonumber \\
	&\equiv \sum_{\alpha = e, m} Q^\alpha_{EM} 
\end{align}
has the meaning of the total heat produced by the dissipative
processes per unit volume of the material. 
In Appendix A, we have shown that $Q^\alpha_{EM}$ has the following closed-form:

\begin{equation}\label{108}
	Q^\alpha_{EM} = {1 \over 2} \int\limits_{-\infty}^{\infty}
	\vF_m^{\,\alpha\dagger}\cdot\left[ \omega\uzpp\,^\alpha +
	\p_{\omega}(\omega\uzpp\,^\alpha)
	(i\p_{t})\right]_{_{\omega_0}}\cdot\vF_m^{\,\alpha}\,dt,
	\qquad (\mbox{where}\ \alpha\ \mbox{is}\ e\ \mbox{or}\ m)
\end{equation} 
Now, using the 6-vector notation, Eq.~(\ref{108}) reads
\begin{equation}\label{109}
	Q_{EM} = {1 \over 2} \int\limits_{-\infty}^{\infty} \vbF_m^{\,\dagger}\cdot\ubA_{s}\cdot\vbF_m\, dt,     
\end{equation}
where $\vbF_{m}^{\dagger}$, the Hermitian conjugate of the complex SVA
of the electromagnetic 6-vector $\vbF_{m}$, reads
$ \left[ \vF^{\,e\,*}_{m},\ \vF^{\,m\,*}_{m} \right]^T $.  In the
above equation, the symmetrized 6-dyadic operator $\ubA_{s}$ is
defined as
\begin{gather}\label{110}
	\ubA_{s} = 
	\begin{bmatrix}
		(\omega\ueppp) + \p_\omega(\omega\ueppp)\overleftrightarrow{i\p_t} && \u0 \\
		\u0 && (\omega\umpp) + \p_\omega(\omega\umpp)\overleftrightarrow{i\p_t}
	\end{bmatrix}_{_{\omega_0}},
\end{gather} 
where $\overleftrightarrow{i\p_t} = i(\p_{t} - \p^{\dagger}_{t})/2$ in
which $\p_{t}$ (or $\p^{\dagger}_{t}$) acts on the function on the
right (or left) side. One can see that only $\uzpp$ which is
responsible for the material loss, appears in Eq.~(\ref{108}).

Having the above discusion in mind and the closed-form of $Q_{EM}$, Eq.~(\ref{109}), we are going to calculate the coupling term, $h_{EM}^\Sigma - p_{rad}^\Sigma$, and the heat flux,
$\vq_{rad}^{\,\Sigma}$, by use of the FEFSVA in the framework of EDGF
formalism \cite{60}.

\subsection{Heating rate density}
Based on the form of Eq. (\ref{109}), we may define the heating rate density, $h_{EM}$, as 
\begin{equation}\label{111}
	h_{EM} = {1 \over 2}\left\langle\vbF_{m}^{\,\dagger}\cdot\ubA_{s}\cdot\vbF_{m}\right\rangle,
\end{equation}
where $\langle\ldots\rangle$ means averaging over the statistical ensemble of time-harmonic fields with random amplitudes and phases. Note that $h_{EM}$ defined in this way takes into account the heat energy associated with the oscillations centered at the carrier frequency $\omega_0$. 

Recalling $\vbF_{m}$ from Eq.~(\ref{17}), the heating rate, Eq.~(\ref{111}), takes on the following form
\begin{equation}\label{112}
	h_{EM}={1 \over 2} \left\langle \vbJ_{m}^{\,\dagger}\cdot\underbrace{\hat{\ubG}\,^{\dagger}\cdot\ubA_{s}\cdot\hat{\ubG}}_{{\equiv\,\hat{\ubG}_{A}}}\cdot\vbJ_{m}\right\rangle
\end{equation}
By using the dyadic algebra definition of trace, we can rewrite Eq.~(\ref{112}) as follows
\begin{equation}\label{113}
	h_{EM}={1 \over 2} \Tr\left[\hat{\ubG}_{A}\cdot
	\left\langle\vbJ_{m}\vbJ_{m}^{\,\dagger}\right\rangle \right].
\end{equation}
In order to calculate the correlation dyadic between the 6-vector fluctuating currents which are SVA, we apply Eq.~(\ref{f_sva}) for the vectors and obtain
\begin{align}\label{114}
	&\left\langle \vbJ_{m}(\vr^{\,\pr},t^{\pr})\vbJ_{m}^{\,\dagger}(\vr^{\:\pr\pr}, t'') \right\rangle =  \nonumber \\ 
	&{1 \over \pi^{2}}\int_{-\Dom/2}^{+\Dom/2}d\Omega'd\Omega''\left\langle \vbJ_{\omega}(\vr^{\,\pr},\omega_{0}+\Omega')\vbJ_{\omega}^{\,\dagger}(\vr^{\:\pr\pr}, \omega_{0}+\Omega'')\right\rangle e^{-i\Omega^{\pr}t'+i\Omega''t''}.
\end{align}

We assume that any physically small volume of the material is at local thermal equilibrium with the current fluctuations obeying the FDT. According to the FDT, Eq.~(\ref{FDT-0}), the correlation dyadic between the time-harmonic 6-vector fluctuating currents with SVA reads
\begin{align}\label{115}
	\bigg\langle\vbJ_{\omega}(\vr^{\,\pr},\omega_{0}+\Omega^{\pr}) & \vbJ_{\omega}^{\,\dagger} (\vr^{\:\pr\pr}, \omega_{0}+\Omega^{\,\pr\pr}) \bigg\rangle = \nonumber \\
	&{4\Theta (\omega_{0}+\Omega^{\pr},T) \over \pi} \ubA_{s,\omega}|_{_{\omega_{0} + \Omega^{\pr}}}\,\delta(\Omega^{\pr}-\Omega^{\pr\pr})\delta(\vr^{\,\prime}-\vr^{\,\prime\prime}),
\end{align}
where $\ubA_{s,\omega}$ is the Fourier transform of Eq. (\ref{110}). 
In Eq. (\ref{115}), the Dirac delta functions indicate spatial or spectral incoherence, respectively. It should be noted that here we assume that there is no correlation between fluctuating electric and magnetic currents, which means that loss mechanisms in the dynamic electric and magnetic responses of the material are independent.

Now, remembering the definition of SVA fields and currents,
Eq.~(\ref{f_sva}), and expanding $\Theta (\omega_{0},T)$ and $\ubA\,^{\alpha}_{s,\omega}({\omega_{0}+\Omega})$ around $\Omega = 0$ while ignoring $O(\Omega^{2})$ and higher terms, after some manipulation, we obtain
\begin{align}\label{116}
	\left\langle \vJ^{\alpha}_{m}(\vr^{\,'},t')\vJ_{m}^{\,\alpha\,\dagger}(\vr^{\,''}, t'') \right\rangle &= {4 \over \pi^{3}}
	\bigg[(\omega\Theta\uzpp_\alpha)_{_{\omega_0}}\int_{-\Dom/2}^{\Dom/2}e^{-i\Omega(t'-t'')}d\Omega' \nonumber \\
	&+\, \p_{\omega}(\omega\Theta\uzpp_\alpha)_{_{\omega_0}}  
	\int_{-\Dom/2}^{\Dom/2}\Omega' e^{-i\Omega(t'-t'')}d\Omega'\, \bigg] ,
\end{align}
where the first integral gives $\Dom[j_{0}(\Dom (t'-t'')/2)]$ (where
$j_0(x) \equiv \sin(x)/x$) and the second integral is just the derivative of the first integral with respect to $t'$ multiplied by $+i$, so that for the 6-vector form we obtain
\begin{align}\label{JJ}
	\left\langle \vbJ_{m}(\vr^{\;\pr},t^{\pr})\vbJ_{m}^{\,\dagger}(\vr^{\;\pr\pr}, t^{\,\pr\pr}) \right\rangle &= \nonumber \\
	&{4 \Dom\over \pi^{3}}\,\ubB_{s}\,j_{0}(\tau''') \delta(\vr^{\;\pr}-\vr^{\;\pr\pr}), \qquad \tau'''=\Dom (t'-t'')/2
\end{align}	 
where the symmetrized operator $\ubB_{s}$ is given by
\begin{gather}\label{118}
	\ubB_{s} = 
	\begin{bmatrix}
		\left( \omega\uep\,^{\pr\pr}\Theta(\omega, T)\right)  + \p_\omega\left(\omega\uep\,^{\pr\pr}\Theta(\omega, T)\right)\overleftrightarrow{i\p_{t^{\pr}}} && \u0 \\
		\u0 && \left( \omega\um\,^{\pr\pr}\Theta(\omega, T)\right)  + \p_\omega \left(\omega \um\,^{\pr\pr} \Theta(\omega, T)\right)  \overleftrightarrow{i\p_{t^{\pr}}}
	\end{bmatrix}_{_{\omega_0}}.
\end{gather}

Recalling Eq.~(\ref{17}) and including Eq.~(\ref{JJ}) in Eq.~(\ref{113}), the heating rate gets the following form
\begin{align}\label{hrat}
	h_{EM}(\vr,t;\omega_0)&={2\Dom\over \pi^{3}} \int d^{3}r'dt'dt'' \nonumber \\
	&\times\Tr\left[\ubG\,^{\dagger}(\vr,t;\vr^{\,\pr}, t'') \cdot \ubA_{s}(\vr,t) \cdot \ubG(\vr,t;\vr^{\;\pr},t') \cdot 
	\ubB_{s}(\vr^{\;\pr},t') \right]_{_{\omega_0}}j_{_0}(\tau''').
\end{align}

\subsection{Radiative power}
Recalling Eq.~(\ref{5-src}) and remembering that the radiative power, $p_{rad}$, is the ensemble average of the source power, $\left\langle P_{src}\right\rangle_T$, we have
\begin{equation}\label{120}
	p_{rad} = - {1 \over 2} \Re\left\langle\vF^{\,e}_{m}\cdot\vJ_{m}^{\,e\,*} + \vF^{\,m\,*}_{m}\cdot \vJ_{m}^{\,m}\right\rangle,
\end{equation} 
where the electric (magnetic) field and current are related by expanding Eq.~(\ref{17}) as follows
\begin{equation}\label{121}
	\begin{bmatrix}
		\vF^{\,e}_{m}\\
		\vF^{\,m}_{m}
	\end{bmatrix} =
	\begin{bmatrix}
		\hat{\uG}_{ee} && \hat{\uG}_{em} \\
		\hat{\uG}_{me} && \hat{\uG}_{mm} 
	\end{bmatrix}\cdot
	\begin{bmatrix}
		\vJ^{e}_{m}\\
		\vJ^{m}_{m}
	\end{bmatrix},
\end{equation}
where $\hat{\uG}_{ee}$, etc., are the elements of $\hat{\ubG}$,	understood as given by Eq. (\ref{18}).
Similar to the heating rate calculation, using Eq.~(\ref{18}) and
Eqs.~(\ref{114})--(\ref{118}) for the electromagnetic fields and
currents vectors with SVA, the radiative power of the fluctuating
fields, $p_{rad}$, Eq.~(\ref{120}), can be written as
\begin{align}\label{123}
	p_{rad}(\vr,t,\omega_{0}) = - {1 \over 2} \Re\!\int d^3r'dt' \Tr\bigg[&\uG_{ee}(\vr,t;\vr^{\;\pr},t')\cdot\left\langle\vJ_{m}^{\,e}(\vr^{\;\pr},t')\vJ_{m}^{\,e^*}(\vr,t)\right\rangle \nonumber \\
	& +\, \uG\,^\dagger_{mm}(\vr,t;\vr^{\,\pr},t')\cdot\left\langle\vJ_{m}^{\,m}(\vr,t)\vJ_{m}^{\,m^*}(\vr^{\,\pr},t')\right\rangle\bigg]_{_{\omega_0}}.
\end{align}
Regarding FDT, we have
\begin{equation}\label{123-extra}
	\left\langle\vJ_{m}^{\alpha}(\vr,t)\vJ_{m}^{\alpha^*}(\vr^{\,\pr},t')\right\rangle
	=
	{4\Dom\over\pi^3}\uB_{s,\alpha}(\vr^{\,\pr},t')j_{_0}(\tau')\delta(\vr-\vr^{\,\pr}),
	\qquad (\alpha\ \mbox{is}\ e\ \mbox{or}\ m),
\end{equation}
where $\tau'=\Dom(t-t')/2$ and $\uB_{s,\alpha}$ is
\begin{equation}\label{124}
	\uB_{s,\alpha}=\left[ \omega\uz\,^{\pr\pr}_{\alpha}\Theta(\omega, T)\right]  + \p_\omega\left[\omega\uz\,^{\pr\pr}_{\alpha}\Theta(\omega, T)\right]\overleftrightarrow{i\p_{t^{\pr}}}. 
\end{equation}
Thus, the closed form of $p_{rad}$ reads
\begin{align}\label{prad}
	p_{rad}(\vr,t,\omega_{0}) &= - {2\Dom \over \pi^3} \Re\!\int dt' \nonumber\\
	&\times\Tr\left[\uG_{ee}(\vr,t,t')\cdot\uB_{s,\,e}(\vr,t') +  	\uG\,^\dagger_{mm}(\vr,t,t')\cdot\uB_{s,\,m}(\vr,t')\right]_{_{\omega_0}}j_{_0}(\tau''').
\end{align}

\subsection{Radiative heat flux}
In order to obtain the closed-form of radiative heat flux, we calculate the ensemble average of Eq.~(\ref{5-poynt}), in the frameworks of the Poynting theorem and the SVA FEF definition. According to $\vq_{rad}$, we have
\begin{equation}\label{126}
	\vq_{rad} = {1 \over 2} \Re\!\left\langle\vF^{\,e}_m\times\vF_m^{\,m\,\dagger}\right\rangle,
\end{equation}
therefore, for the radiative heat flux component perpendicular to the interface, $q_{rad,n}$, we have
\begin{equation}\label{127}
	q_{rad,n} = {1 \over 2} \Re\!\left\langle\epsilon_{nkl}F^{\,e,\,k}_m F^{\,m,\,l\,\dagger}_m\right\rangle,
\end{equation}
where $\epsilon_{nkl}$ is the Levi-Civita symbol and the summation over repeating (dummy) indices is implied (also known as Einstein's summation notation). 
Recalling Eqs.~(\ref{121}) and (\ref{18}), we can write
\begin{align}\label{128}
	q_{rad,n} = {1 \over 2} \Re\bigg\langle & \epsilon_{nkl}\int d^3r'd^3r''dt'dt'' \nonumber \\
	&\times \hat{e}_k\cdot\left[ \uG_{ee}\cdot\vJ^e_{m} + \uG_{em}\cdot\vJ^m_{m}\right] 
	\left[ \uG_{me}\cdot\vJ^e_{m} + \uG_{mm}\cdot\vJ^m_{m}\right]^\dagger\cdot\hat{e}_l\bigg\rangle_{_{\omega_0}}.
\end{align}
Given the absence of correlation between the electric and magnetic components of fluctuating electromagnetic currents, 
we can derive the electric/magnetic component of $q_{rad,n}$ as follows:
\begin{equation}\label{129}
	\begin{aligned}
		q^\alpha_{rad,n} &= {1 \over 2}\epsilon_{nkl} \Re\! \int d^3r'd^3r''dt'dt'' \\
		&\times \bigg[\hat{e}_k\cdot\uG_{e\alpha}(\vr,t;\vr^{\;\pr},t')\cdot
		\bigg\langle\vJ^\alpha_{m}(\vr^{\;\pr},t')\vJ^{\alpha\,\dagger}_{m}(\vr^{\;\pr\pr},t'')\bigg\rangle
		\cdot\uG\,^\dagger_{m\alpha}(\vr,t;\vr^{\;\pr\pr},t'')\cdot\hat{e}_l\bigg]_{_{\omega_0}}.  
	\end{aligned}
\end{equation}
Using the SVA approach for fluctuating electric/magnetic vector
currents and then imposing FDT, (cf. Eqs.~(\ref{114})--(\ref{118})),
to calculate the correlation between them, i.e.,
$\left\langle\vJ^\alpha_{m}(\vr^{\,\pr},t')\vJ^{\alpha\,\dagger}_{m}(\vr^{\,\pr\pr},t'')\right\rangle
$, we obtain the radiative heat flux expressed in a closed form as follows
\begin{equation}\label{qrad-1}
	\begin{aligned}
		q_{rad,n}(\vr,t;\omega_{0}) &= {2\Dom \over \pi^3}\epsilon_{nkl} \hat{e}_k\cdot\bigg[\Re\!\!\sum_{\alpha=e,m}\!
		\int d^3r'dt'dt'' \\
		&\times \uG_{e\alpha}(\vr,t;\vr^{\,\pr},t')\cdot \uB_{s,\alpha}(t')\cdot
		\uG\,^{\dagger}_{m\alpha}(\vr,t;\vr^{\,\pr},t'')j_{_0}(\tau''')\bigg]_{_{\omega_0}}\!\cdot\hat{e}_l,
	\end{aligned}
\end{equation}
which, in terms of the EDGF components, can be written by using Einstein's summation notation as follows:
\begin{equation}\label{qrad-2}
	\begin{aligned}
		q_{rad,n}(\vr,t;\omega_{0}) &= {2\Dom \over \pi^3}\epsilon_{nkl} \Re\! \int d^3r'dt'dt'' \\
		&\times \sum_{\alpha=e,m}\bigg[G^{k\lambda}_{e\alpha}(\vr,t;\vr^{\,\pr},t')B^{\lambda\sigma}_{s,\alpha}(t')
		G^{\sigma l\dagger}_{m\alpha}(\vr,t;\vr^{\,\pr},t'')\bigg]_{_{\omega_0}}j_{_0}(\tau''').
	\end{aligned}
\end{equation}

Having the closed form relations of the heating rate, radiative power,
and radiative heat flux for any DDANM in which we can apply our EDGF
formalism \cite{60} rather than the standard Green's function method,
we are now ready to employ our DHT model.  In the next section, as a
suitable test problem, we consider the case of paraxial heat transfer
DDANM. The paraxial treatment is also applicable for the RHT in
uniaxial crystals of dielectric nanolayers or plasmonic nanorods.

\section{DHT in paraxial media}
In this section, we model the dynamics of RHT in uniaxial DDANM in
which $|\epz| \gg |\ept|$ and/or $|\mz| \gg |\mt|$. In this media, the
radiative heat propagation happens dominantly along the anisotropy
axis. Thus, such media can be called paraxial media. In order to model
DHT in paraxial media, we need some tools to obtain the paraxial EDGF
(PEDGF), which will be employed for the relevant calculations in this
section. Appendix B summarizes such necessary tools of the PEDGF
approach.

The components of the PEDGF, Eq.~(\ref{PEDGF}), are given as
follows:
\begin{align}\label{Comp-PEDGF1}
	\uG\,^{a,\gamma}_{\alpha\beta} = G\,^{a,\gamma}_{\alpha\beta}(\tau'_{a})\,\uI\,^{\gamma}_{\alpha\beta}(\vec{\varrho},|Z|),
\end{align}
\begin{align}\label{Comp-PEDGF2}
	G\,^{a,\gamma}_{\alpha\beta} (\tau'_{a})= {\Dom \over (2\pi)^{3}}\, g^{a,\gamma}_{\alpha\beta}(\tau'_{a}), \qquad  \tau'_{a}=\tau_{a}\Dom/2 ,
\end{align}
\begin{align}\label{Comp-PEDGF3}
	g^{a,\gamma}_{\alpha\beta}(\tau'_{a}) = C\,^{a,\gamma}_{\alpha\beta} j_{0}(\tau'_{a}) - 
	i(\Dom/2) D\,^{a,\gamma}_{\alpha\beta} j_{1}(\tau'_{a}),
\end{align}		
\begin{align}\label{Comp-PEDGF4}
	\uI\,^{\gamma}_{\alpha\beta}(\vec{\varrho},|Z|) = e^{i\kappa_a|Z|}\int d\vk_{t} e^{i[\vk_{t}\cdot\vec{\varrho}-\chi^{\gamma}_{a}k_t^2]}\,\ue\,^{\gamma}_{\alpha\beta}.
\end{align}
The meanings of all quantities, variables, and indices in
Eqs.~(\ref{Comp-PEDGF1})-(\ref{Comp-PEDGF4}) are explained in Appendix
A, however, since superscript and subscript notations play an
important role in the rest of this paper, we recall them as follows:

$\textit{a}$: denotes the paraxial approach; 

$\gamma$: labels two orthogonal polarizations, that is, $p$ (TE) and $s$ (TM); 

$\alpha$ and $\beta$: signifies the electric and magnetic components of the quantity, that is, $e$ or $m$, such that $\alpha\alpha$ ($\beta\beta$) means $ee$ ($mm$) and $\alpha\beta$ ($\beta\alpha$) means $em$ ($me$).

Before calculating the heating rate, radiative heat power, and radiative heat flux within the framework of PEDGF approach, we consider two following assumptions to find a closed-form relation for paraxial heating rate (PHR):

\textit{First}, we assume that the time variation of temperature is so slow with respect to time variation of $j_{_0}(\tau''')$ so that we can write
\begin{equation}\label{assum-1}
	\Theta(\omega,T(t'))\approx \Theta(\omega,T(t'')), \qquad t\neq t''
\end{equation}
Of course, it does not mean that $\p_{t'}T(t')=0$ and we can consider
$\p_{t'}T$, which is denoted by $\dot{T}'$, as a constant on the time scale of FEFs. With this assumption we can write
\begin{equation}\label{143}
	\p_{t'}b^\alpha_{n_i} = \dot{T}'\p_{T}(b^\alpha_{n_i}), \qquad n_i=0,1  \quad \mbox{and} \quad i = t, z,
\end{equation}
where $t$ and $z$ denote the transverse and axial directions, respectively, and $\p_{T}={\p \over \p_{T}}$. 

\textit{Second}, we consider low frequency or high temperature applications for which we have $\big|\hbar\omega_0/k_BT\big|\ll 1$ so that
\begin{equation}\label{assum-2}
	\Theta(\omega_{0},T) = {\hbar\omega_0 \over e^{\hbar\omega_0/k_BT}-1 } \simeq k_BT.
\end{equation}
Since we assumed that the electromagnetic constitutive parameters of the medium, i.e. $\vep$ and $\mu$, are not time- and temperature-dependent, by use Eqs.~(\ref{assum-2}) and (\ref{143}) we have 
\begin{equation}\label{145}
	b^\alpha_{0_i}=	k_BT a^\alpha_{0_i} \qquad b^\alpha_{1_i} = k_B \dot{T}' a^\alpha_{1_i}.
\end{equation}	

\subsection{Paraxial heating rate}
Given two aforementioned assumptions, Eqs. (\ref{assum-1}) and
(\ref{assum-2}), we aim to obtain the PHR. 
Starting Eq.~(\ref{hrat}), the PHR reads
\begin{equation}\label{phrat}
	\begin{gathered}
		h^a_{EM}(\vr,t;\omega_0,\Dom)={2\Dom\over \pi^{3}} \int d^{3}r'dt'dt''\,\Tr\left[\ubGt\right]_{_{\omega_0}}
		j_{_0}(\tau'''),\\
		\ubGt=\ubG\,^{a\,\dagger}(\vr,t;\vr^{\,\pr}, t'') \cdot \ubA_{s}(\vr,t) \cdot \ubG\,^a(\vr,t;\vr^{\,\pr},t') \cdot 
		\ubB_{s}(\vr^{\,\pr},t'),
	\end{gathered}
\end{equation}
where $\tau'''=\Dom(t'-t'')/2$ and $\ubG\,^a$ is PEDGF, Eq.~(\ref{PEDGF}). 
To do integrating over $t'$ and $t''$, let us calculate the integrand in above equation.  $\Tr[\ubGt]_{_{\omega_0}}j_{_0}(\tau''')$ has the following closed-form representation (see Appendix C):
\begin{align}\label{146}
	&\Tr\left[\ubGt\,^{\gamma}_{\alpha\beta}\right]_{_{\omega_0}} j_{_0}(\tau''')= \nonumber \\
	& k_BT \sum_{\alpha,\beta,\gamma}\bigg[  a^\alpha_{0_t}(a^\beta_{0_t} + a^\beta_{1_t} \dot{T}' / 2T) \left(\uG\,^{a,\gamma\,\dagger}_{\alpha\beta}(t,t''):\uG\,^{a,\gamma}_{\alpha\beta}(t,t')\right)j_{_0}(\tau''') \nonumber \\
	& - {i\Dom\over 4} a^\alpha_{1_t} a^\beta_{0_t} \left(\uH\,^{a,\gamma\,\dagger}_{\alpha\beta}(t,t''):\uG\,^{a,\gamma}_{\alpha\beta}(t,t') + 
	\uG\,^{a,\gamma\,\dagger}_{\alpha\beta}(t,t''):\uH\,^{a,\gamma}_{\alpha\beta}(t,t')
	\right)j_{_0}(\tau''')  \nonumber \\
	& + {i\Dom\over 4} a^\alpha_{0_t} a^\beta_{1_t} \bigg(\uG\,^{a,\gamma\,\dagger}_{\alpha\beta}(t,t''):\uH\,^{a,\gamma}_{\alpha\beta}(t,t')j_{_0}(\tau''') 
	- \uG\,^{a,\gamma\,\dagger}_{\alpha\beta}(t,t''):\uG\,^{a,\gamma}_{\alpha\beta}(t,t')j_{_1}(\tau''')\bigg) \bigg]_{_{\omega_0}}.	\nonumber\\
\end{align}	 

Recalling Eqs.~(\ref{Comp-PEDGF1})-(\ref{Comp-PEDGF4}), the trace of
tensor multiplications between different dyadics in Eq.~(\ref{146})
are reduced with the help of the following expression
\begin{align}\label{intgl-1}
	\uI\,^{\gamma\,\dagger}_{\alpha\beta}:\uI\,^{\gamma}_{\alpha\beta}=\big|e^{i\kappa_{a}|Z|}\big|^2\int d\vk'_{t}d\vk_{t} I^{\gamma *} (\vk'_{t},\vvro) I^\gamma(\vk_{t},\vvro)\left[\ue\,^{\gamma\,\dagger}_{\alpha\beta}(\vk'_t):\ue\,^{\gamma}_{\alpha\beta}(\vk_t)\right], 
\end{align}
where $I^{\gamma}(\vk_{t},\vvro)=e^{i[\vk_{t}\cdot\vvro-\chi^{\gamma}_{a}k_t^2]}$ with $\vvro = \vec{\rho} - \vro^{\,\prime}$.
As explained in Appendix D, the integration over $\vk_{t}$ leads to Eq.~(\ref{uI}) and we have
\begin{equation}\label{intgl-2} 
	\uI\,^{\gamma\,\dagger}_{\alpha\beta}:\uI\,^{\gamma}_{\alpha\beta} = \big|I_\nu^\gamma(\rho', |Z|)\big|^2,
\end{equation}
where $\big|I_\nu^\gamma(\rho', |Z|)\big|^2$ with $\nu= \{ee, em, me,
mm\}$ is given by Eq.~(\ref{uI2}) in which we have assumed that the
observation point takes place on the $z$-axis.
Now, recalling Eqs.~(\ref{Comp-PEDGF1})-(\ref{Comp-PEDGF4}), we obtain 
\begin{align}\label{148}
	\uP\,^{a,\gamma\,\dagger}_{\alpha\beta}:\uP\,^{a,\gamma}_{\alpha\beta} = \mathfrak{P}^{\,a,\gamma\,*}_{\alpha\beta}(\tau_{a}'')\mathfrak{P}^{\,a,\gamma}_{\alpha\beta}&(\tau_{a}') = \nonumber \\
	&\left[{\Dom \over (2\pi)^3} \right]^2 \mathfrak{p}^{\,a,\gamma\,*}_{\alpha\beta}(\tau_{a}'')\mathfrak{p}^{\,a,\gamma}_{\alpha\beta}(\tau_{a}')\big|I_\nu^\gamma(\rho', |Z|)\big|^2,
\end{align}
where $\mathfrak{P}$ and $\mathfrak{p}$ denote any of $\{G,H$\} and $\{g,h\}$, respectively, and $\tau_{a}^{\pr[\pr\pr]}=(t-t^{\pr[\pr\pr]})\Dom/2$.
By substituting Eq.~(\ref{148}) in Eq.~(\ref{146}), we obtain
\begin{align}\label{150}
	\Tr&\left[\ubGt\,^{\gamma}_{\alpha\beta}\right]_{_{\omega_0}}j_{_0}(\tau''')=\left[{\Dom \over (2\pi)^3} \right]^2 k_BT\, \sum_{\alpha,\beta,\gamma} \big|I_\nu^\gamma(\rho', |Z|)\big|^2_{_{\omega_0}} \nonumber \\ 
	& \times \bigg\{ a^\alpha_{0_t}(a^\beta_{0_t} + a^\beta_{1_t} \dot{T}' / 2T) \left[g^{\,a,\gamma\,*}_{\alpha\beta}(\tau_{a}'')g^{\,a,\gamma}_{\alpha\beta}(\tau_{a}')\right] j_{_0}(\tau''') \nonumber \\
	&- {i\Dom\over 4} a^\alpha_{1_t} a^\beta_{0_t} \left[h^{\,a,\gamma\,*}_{\alpha\beta}(\tau_{a}'')g^{\,a,\gamma}_{\alpha\beta}(\tau_a') + g^{\,a,\gamma\,*}_{\alpha\beta}(\tau_{a}'')h^{\,a,\gamma}_{\alpha\beta}(\tau_{a}')\right] j_{_0}(\tau''') \nonumber \\
	&+ {i\Dom\over 4} a^\alpha_{0_t} a^\beta_{1_t}
	\left[g^{\,a,\gamma\,*}_{\alpha\beta}(\tau_{a}'')h^{\,a,\gamma}_{\alpha\beta}(\tau_{a}')j_{_0}(\tau''') - 
	g^{\,a,\gamma\,*}_{\alpha\beta}(\tau_{a}'')g^{\,a,\gamma}_{\alpha\beta}(\tau_{a}')j_{_1}(\tau''')\right] \bigg\}_{_{\omega_0}}. \nonumber \\
\end{align}
Thus, with Eq.~(\ref{150}) in hand, the final form of PHR is given by
\begin{align}\label{fphrat}
	h^{a,c}_{EM}(\vr,t;\omega_0,\Dom)&={2\Dom\over \pi^{3}}{k_B\over (2\pi)^{4}}
	\sum_{\alpha,\beta,\gamma} \int d^3r'\big|I_\nu^\gamma(\rho', |Z|)\big|^2_{_{\omega_0}} T(\vr^{\,\pr}) \nonumber\\
	&\times \bigg[a^\alpha_{0_t}(a^\beta_{0_t} + a^\beta_{1_t} \dot{T} / 2T) [ |C^{\,a,\gamma}_{\alpha\beta}|^2 + 
	\Dom^2 |D^{\,a,\gamma}_{\alpha\beta}|^2/12 ] \nonumber \\
	& - {\Dom^2\over 4} a^\alpha_{0_t} a^\beta_{1_t}
	\Re\left(C^{\,a,\gamma}_{\alpha\beta} \, D^{\,a,\gamma\,*}_{\alpha\beta}\right)\bigg]_{_{\omega_0}},
\end{align}
where the superscript $c$ denotes that we work in the framework of slow time variation of the temperature and also focus on the low frequency/high temperature applications.

\subsection{Uniaxial radiative power}
Given the aforementioned assumptions (Eqs.~(\ref{assum-1}) and (\ref{assum-2})), we present a closed-form relation for the uniaxial radiative power. 
It should be noticed, as explained below, that here we calculate
$p_{rad}$ in a more general case than what is given by the paraxial
approximation, that is, here we are interested in the power density of
thermal radiation in a unixial medium.  Including PEDGF, $\ubG\,^a(\vr,t,t')$, in the $p_{rad}$ relation, Eq.~(\ref{prad}), and going forward similarly to the PHR case, we need to calculate the integral $I_\gamma(\vvro, |Z|)$ at $\vvro=0$ and
$|Z|=0$, which causes divergency problem (see Eq.~(\ref{uI}) in Appendix D).

In order to model uniaxial radiative power, we start from Eq.~(\ref{prad}). 
By expanding two terms in the bracket of Eq.~(\ref{prad}) as follows
\begin{align}\label{154-extra}
	\Tr\left[ \ubG_{ee}\cdot\uB_{s,e}\right]_{_{\omega_0}}\!j_{_0}(\tau') &= \nonumber \\
	&\bigg[ \uG\,^{tt,p}_{ee}(\vr,t,t'):\uB\,^{t}_{s,e}(t') + \uG\,^{tt,s}_{ee}(\vr,t,t'):\uB\,^{t}_{s,e}(t') \nonumber \\ 
	&+ \uG\,^{zz,p}_{ee}(\vr,t,t'):\uB\,^{z}_{s,e}(t')\bigg]_{_{\omega_0}}\!j_{_0}(\tau'), \nonumber \\
	\Tr\left[ \ubG\,^{\dagger}_{mm}\cdot\uB_{s,m}\right]_{_{\omega_0}}\!j_{_0}(\tau') &= \nonumber \\
	&\bigg[ \uG\,^{tt,p\,\dagger}_{mm}(\vr,t,t'):\uB\,^{t}_{s,m}(t') + \uG\,^{tt,s\,\dagger}_{mm}(\vr,t,t'):\uB\,^{t}_{s,m}(t') \nonumber \\
	&+ \uG\,^{zz,s\,\dagger}_{mm}(\vr,t,t'):\uB\,^{z}_{s,m}(t')\bigg]_{_{\omega_0}}\!j_{_0}(\tau'),
\end{align}

the electric (magnetic) component of uniaxial radiative power can be written as (see Appendix E)
\begin{align}\label{158}
	p^{e[m]}_{rad} &=+[-]{\Dom k_B T \over \pi^4} \Re\!\sum_{j,\gamma} \!\int\! k_{t}dk_{t} \left(1-\delta_{jz}\delta_{\gamma,s[p]}\right) \nonumber \\
	&\times \left[ \left( a^{e[m]}_{0_j}+ia^{e[m]}_{1_j}\dot{T}/2T\right) C^{jj,\gamma[\,*]}_{ee[mm]}+{\Dom^2 \over 12} a^{e[m]}_{1_j } D^{jj,\gamma[\,*]}_{ee[mm]} \right]_{_{\omega_0}} .
\end{align}
Now, with $C$'s and $D$'s in hand (from $\ubC$ and $\ubD$ in \cite{60}), we can integrate over $\vk_{t}$. It should be noted that $|\vk_{t}|$ varies from $0$ to $k^\gamma_m$ where $k^\gamma_m$ is calculated by considering the maximum value of $\kappa^\gamma_{0}$ for the \textit{low loss media}. So, we have $k^{\,p[s]}_m \approx \sqrt{a^{m[e]}_ta^{e[m]}_z}$. This assumption is reasonable if we remember that here the propagating fields are needed to be considered rather than the evanescent ones. 
So, by this assumption and after integrating over $\vk_{t}$, we obtain
\begin{align}\label{fprad}
	p^{c}_{rad} &= p^{e}_{rad} + p^{m}_{rad}, \nonumber\\
	p^{e[m]}_{rad} &=-[+]{\Dom \over \pi^4} k_B T\, \Re\!\sum_{\gamma,j}\left[\Pi^{\gamma[*]}_{e[m],j}\, f^{e[m]}_j(T) + {\Dom^2 \over 12} \p_{\omega}(\Pi^{\gamma[*]}_{e[m],j}) a^{e[m]}_{1_j}\right]_{_{\omega_0}},
\end{align}	
where 	
\begin{align}\label{fprad-coeff}
	&f^{e[m]}_j(T) =a^{e[m]}_{0_j}+ia^{e[m]}_{1_j}\dot{T}/2T, \nonumber \\ 
	&\Pi_{e[m],t}^{p[s]} = {a^{m[e]}_t\kappa_{a}\over 6}{a^{e[m]}_z \over a^{e[m]}_t}, \nonumber \\ 
	&\Pi_{e[m],t}^{s[p]} = {a^{m[e]}_z\kappa_{a}\over 2}, \nonumber \\ 
	&\Pi_{e[m],z}^{p[s]} = {a^{m[e]}_t\kappa_{a}\over 3}, \nonumber  \\ 
	&\Pi_{e[m],z}^{s[p]} = 0.
\end{align}
In Eq.~(\ref{fprad}), the superscript $c$ denotes that we work in the framework of the slow time variation of temperature and also focus on the low frequency/high temperature applications.
It should be emphasized that in order to calculate
$\p_{\omega}\Pi^{\gamma}_{em,tz}$ the expressions for $a^{e,m}_{t,z}$
and $\kappa_{a}$ should be considered as functions of $\omega$ before
taking their values at the center frequency $\omega_{0}$.

\subsection{Paraxial radiative flux}
	In this subsection, we present a closed-form of the paraxial radiative heat flux given the aforementioned assumptions formulated by Eqs.~(\ref{assum-1}) and (\ref{assum-2}). 
Recalling Eq.~(\ref{qrad-2}) and remembering that $\uB_{s,\alpha}$ has a block-diagonal matrix representation, i.e., $\uB_{s,\alpha}=B^\alpha_{s,t}\uI_{t}+B^\alpha_{s,z}\uI_{z}$ for reciprocal media (cf. Eqs.~(\ref{141-1})-(\ref{141-2})), we have $B^{\lambda\sigma}_{s,\alpha}=0$ when
$\lambda\ne\sigma$. We follow the notation from paraxial section and
keep $\alpha=e,m$. Recalling the components of PEDGFs and keeping just
$G^{tt}$ (which is the dominant component in the paraxial case) in Eq.~(\ref{qrad-1}), $q_{rad}$ reads
\begin{equation}\label{159}	
	q_{rad,n}(\vr,t;\omega_{0}) = {2\Dom \over \pi^3}\epsilon_{nkl} \hat{e}_k\cdot\left[\Re\!\sum_{\alpha=e,m}\!
	\int d^3r'dt'dt''\,\ucG\,^{a}_{e\alpha} j_{_0}(\tau''')\right]_{_{\omega_0}}\!\cdot\hat{e}_l,
\end{equation}
where $\ucG\,^{a}_{e\alpha}$ is given by
\begin{equation}\label{159-1}
	\ucG\,^{a}_{e\alpha}j_{_0}(\tau''')=\sum_{\gamma=p,s}\uG\,^{a,\,\gamma}_{e\alpha}(\vr,t;\vr^{\,\pr},t')\cdot \uB_{s,\alpha}(t')\cdot \uG\,^{a,\,\gamma\dagger}_{m\alpha}(\vr,t;\vr^{\,\pr\pr},t'')j_{_0}(\tau''').
\end{equation}
In Appendix F, we have obtained a closed-form formula for paraxial radiative flux in the direction $n \equiv z$  which takes on the following form:
\begin{align}\label{166}	
	q_{rad,z}(\vr,t;\omega_{0}) = {\Dom \over \pi^4}{k_B T \over (2\pi)^3} &\Re\!\sum_{\alpha}
	\int d^3r'\,\big|I_\nu^p(\rho', |Z|)\big|^2 \nonumber \\
	&\times \bigg[ (a^\alpha_{0_t}+a^\alpha_{1_t}\dot{T}/2T) ( C^{a,p}_{e\alpha} C^{a,p\,*}_{m\alpha} + {\Dom^2 \over 12} D^{a,p}_{e\alpha} D^{a,p\,*}_{m\alpha} ) \nonumber \\
	& - {\Dom^2 \over 12} a^{\alpha}_{1_t} \left( C^{a,p}_{e\alpha} D^{a,p\,*}_{m\alpha}
	+ D^{a,p}_{e\alpha} C^{a,p\,*}_{m\alpha} \right) \bigg]_{_{\omega_0}}. 
\end{align}	
Regarding Eq.~(\ref{pCD's}), $C^p_{em}$ and $D^p_{em,me}$ are zeros and thus, $q_{rad,z}$ reads
\begin{align}\label{fqrad}	
	q^c_{rad,z}(\vr,t;\omega_{0}) &= {\Dom \over 2\pi^4}{k_B T \over (2\pi)^3} n_z \nonumber \\
	&\times \Re\!\bigg[
	\int d^3r'\,\big|I_\nu^p(\rho', |Z|)\big|^2\,
	{\Dom^2 \over 12} a^{\alpha}_{1_t} D^{a,p}_{ee}
	-(a^\alpha_{0_t}+a^\alpha_{1_t}\dot{T}/2T) C^{a,p}_{ee} \bigg]_{_{\omega_0}},
\end{align}	
where $n_z=\mbox{sgn}(z-z')$.

We still need to integrate over the space coordinates, which requires
knowing the spatial dependency of the temperature (see the next section).

\section{Temperature profile of paraxial reciprocal media}

\begin{figure}[h]
	\centering
	\includegraphics[scale=.5]{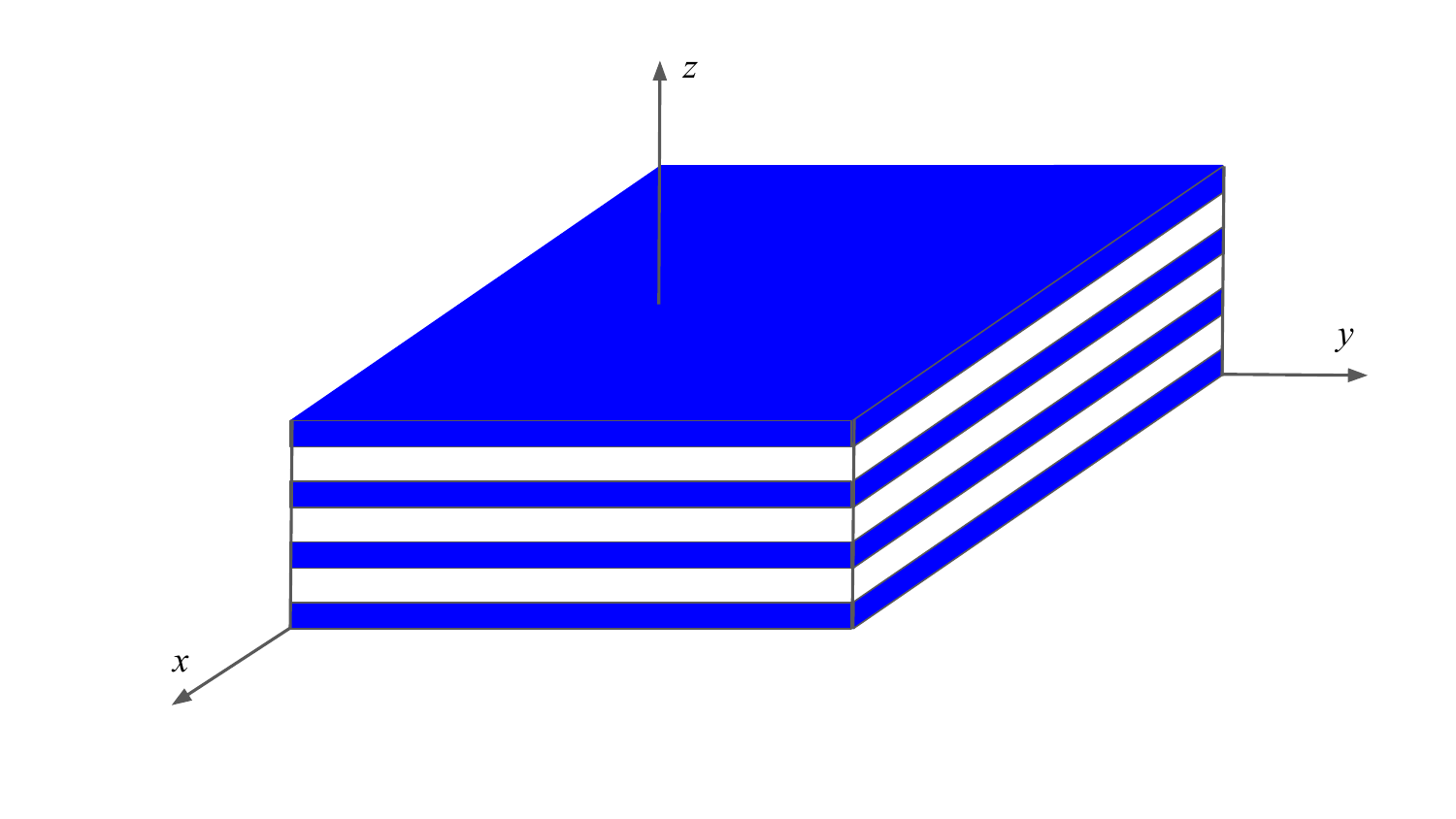}
	\caption{An illustration of a uniaxial nanolayered metamaterial}
	\label{2ndfig}
\end{figure}

Now, for the sake of completeness, we shall find a matrix equation to
find the temperature profile of a layered uniaxial medium
(Fig.~\ref{2ndfig}) which has been made by repeated copies of a
bilayer reciprocal medium for which we can define the effective transverse and axial
phononic thermal conductivities, based on the effective medium theory,
as $\ua_{c}=\ua_{c_t}+\ua_{c_z}$. For simplicity and without losing
the generality, we assume $g_c=0$ and $dL_p/dt=0$ in
Eq.~(\ref{3-HDE}). Working in the cylindrical coordinate system, it is
also a reasonable assumption for paraxial media that there is no
azimuthal variation in temperature profile. Now, by rewriting
Eq.~(\ref{3-HDE}) in the cylindrical coordinate system we have
\begin{equation}\label{HDE-0}
	\bar{\alpha}_{c_v}\dot{T}(\rho,z,t) + \alpha_{c_t}\left(
	\rho^{-1}\p_{\rho} + \p^2_{\rho}\right) T(\rho,z,t) +
	\alpha_{c_z}\p^2_z T(\rho,z,t) = \mathcal{S}_{hp}(\rho,z,t), 
\end{equation}
where $\mathcal{S}_{hp} = (h_{EM}-p_{rad})_{_{\omega_0}}$ and
$\bar{\alpha}_{c_v}$ is the volumetric averaged value of
$\varrho_{m}c_v$ for the medium. In the above equation, within the framework of \textit{effective medium theory} \cite{EMT}, $\alpha_{c_z}$ and $\alpha_{c_t}$ denote
\begin{equation}\label{171}
	\alpha_{c_z} = {\alpha^m_{c}\alpha^h_{c}\over \alpha^m_{c}(1-f) + \alpha^h_{c}f}, \qquad 
	\alpha_{c_t} = \alpha^m_{c} f + \alpha^h_{c}(1-f),
\end{equation}
with $\alpha^{m(h)}_{c}$ and $f$ as the phononic thermal conductivity
of the embedded medium (or host) and the volume fraction of the medium, respectively. Recalling Eqs.~(\ref{assum-1}) and (\ref{assum-2}), the right-hand side of Eq.~(\ref{HDE-0}), $\mathcal{S}_{hp}(\vr,t)$ is replaced by $\mathcal{S}^c_{hp}$ which is defined as $\mathcal{S}^{c}_{hp} = (h^{a,c}_{EM}-p^{c}_{rad})_{_{\omega_0}}$ (cf. Eqs.~(\ref{fphrat}) and (\ref{fprad})).

In order to obtain the profile of temperature, the non-homogeneous integro-differential equation, Eq.~(\ref{HDE-0}), needs to be solved.
We aim to present a closed-form relation for solution of
Eq.~(\ref{HDE-0}). For this purpose, we consider the \textit{stationary state} heat transfer in the medium. By setting $\dot{T}=0$ in Eqs.~(\ref{fphrat}) and (\ref{fprad}), Eq.~(\ref{HDE-0}), in the cylindrical coordinates, reads
\begin{equation}\label{HDE-1}
	\alpha_{c_t}\nabla_\rho^2 T(\rho,z) + \alpha_{c_z}\p^2_z T(\rho,z) = \mathcal{S}^{c,st}_{hp}(\rho,z), 
\end{equation}
where the right-hand side of the equation above is given by
\begin{equation}\label{st-0}
	\mathcal{S}^{c,st}_{hp} = (h^{a,c,st}_{EM}-p^{c,st}_{rad})_{_{\omega_0}},
\end{equation}
In Eq.~(\ref{st-0}), the superscript $st$ emphasizes that we considered steady state conditions and $h^{a,c,st}_{EM}$ and $p^{c,st}_{rad}$ call
\begin{align}\label{hp-st}
	h^{a,c,st}_{EM} =& k_B \sum_{\alpha,\beta,\gamma} F^{a,c, st}_{\alpha\beta,\gamma} \int_{z_i}^{z_f} dz' \int_{0}^{\infty}\rho' d\rho' |I^{\gamma}_{\nu}(\rho', |z-z'|)|^2 T(\rho',z'), \nonumber \\
	p^{c,st}_{rad} =& k_B \sum_{\alpha,\gamma} Q^{c,st}_{\alpha,\gamma} T(\rho,z),
\end{align}
with $F^{a,c,st}_{\alpha\beta,\gamma}$ and $Q^{c,st}_{\alpha,\gamma}$
expressed as follows	
\begin{align}\label{FQ}
	F^{a,c, st}_{\alpha\beta,\gamma} = {\Dom\over \pi^{4}} &\bigg[a^\alpha_{0_t}a^\beta_{0_t} (|C^{\,a,\gamma}_{\alpha\beta}|^2 + 
	\Dom^2 |D^{\,a,\gamma}_{\alpha\beta}|^2/12 ) - {\Dom^2\over 4} a^\alpha_{0_t} a^\beta_{1_t}
	\Re\!\left(C^{\,a,\gamma}_{\alpha\beta} \, D^{\,a,\gamma\,*}_{\alpha\beta}\right)\bigg]_{_{\omega_0}}, \nonumber \\
	&Q^{c,st}_{e[m],\gamma} =-[+]{\Dom \over \pi^4} \Re\!\sum_{j}\left[\Pi^{\gamma[*]}_{e[m],j}\, a^{e[m]}_{0j} + {\Dom^2 \over 12} \p_{\omega}(\Pi^{\gamma[*]}_{e[m],j}) a^{e[m]}_{1_j}\right]_{_{\omega_0}}.
\end{align}
In order to solve Eq.~(\ref{HDE-1}), let us rewrite it as follows
\begin{align}\label{HDE-1-extra0}
	&\alpha_{c_t}\left( \rho^{-1}\p_{\rho} + \p^2_{\rho}\right)T(\rho,z) + \alpha_{c_z}\p^2_z T(\rho,z) + \alpha_Q T(\rho,z) = \mathcal{T}^{a,c, st}, \nonumber \\
	&\mathcal{T}^{a,c, st} = k_B \sum_{\alpha,\beta,\gamma, \omega_0} F^{a,c, st}_{\alpha\beta,\gamma} \int_{z_i}^{z_f} dz' \int_{0}^{\infty}\rho' d\rho' |I^{\gamma}_{\nu}(\rho', |z-z'|)|^2 T(\rho',z'),
\end{align}
where $\alpha_Q = k_B \sum_{\alpha,\gamma, \omega_0} Q^{c,st}_{\alpha,\gamma}$.
By setting a Gaussian transverse profile, we consider the following expansion of $T(\rho,z)$ in the cylindrical coordinates 
\begin{equation}\label{HDE-1-extra2}
	T(\rho, z) = e^{-\alpha_\sigma\rho^2}\sum_{n=0}^{\infty}\rho^n\,T_n(z), \qquad \alpha_\sigma = 1/(2\sigma_{\rho})\,.
\end{equation}
Thus, to find the temperature profile, we need to obtain $T_n(z)$ for
each value of the order $n=0,1,2, ...$ at each point $z \in (z_i,
z_f)$, i.e. between the input and output interfaces of a nanolayered metamaterial, respectively. 

In order to obtain $T_n(z)$, first, we ignore the source term in Eq.~(\ref{HDE-1-extra0}), $\mathcal{T}^{a,c, st}$, and thus we obtain the following differential equation:
\begin{equation}\label{HDE-1-extra1}
	\alpha_{c_t}\left( \rho^{-1}\p_{\rho}+\p^2_{\rho}\right)T(\rho,z) + \alpha_{c_z}\p^2_z T(\rho,z) 
	= 0.
\end{equation}
Next, we recall the Gaussian transverse temperature profile and the
solution presented by Eq.~(\ref{HDE-1-extra2}), and, by employing an
analogy with the method of separation of variables and using the
Frobenius method within the yielded differential equation, we look for a
second-order differential equation for $T_n(z)$. Finally, we include
the source term and find an integro-differential equation that
has to be solved at each value of $n$.

\begin{figure}[h]
	\centering
	\includegraphics[scale=.65]{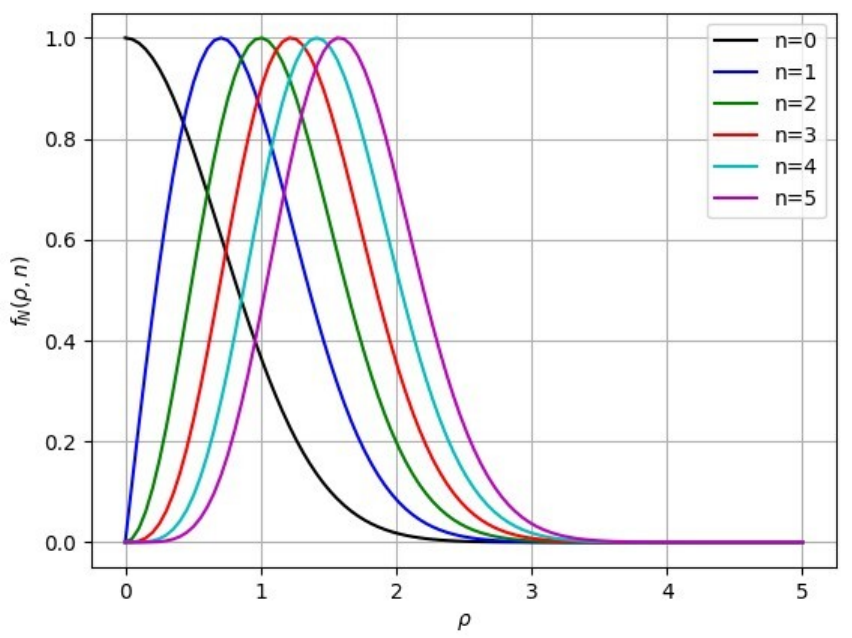}
	\caption{ Variation of $f_\mathcal{N}(\rho, n)$ versus $\rho$ for $n=0-5$ and $\alpha=1$ (see text).}
	\label{3rdfig}
\end{figure}

Before implementing this strategy, let us take a closer look at $f(\rho, n)=\rho^n e^{-\alpha_\sigma\rho^2}$.
In Fig.~\ref{3rdfig}, we plot $f_\mathcal{N}(\rho, n)$ versus $\rho$ for
$n=0,1,\ldots 5$, where the index $\mathcal{N}$ denotes that $f(\rho, n)$
has been normalized by its maximum value at each $n$. In this
numerical example, we set $\sigma_{\rho} = 0.5$. As can be seen, the
peak of $f_\mathcal{N}(\rho, n)$ shifts forward as $n$ increases and the peak
positions become closer to each other at higher $n$'s.

In order to find $T_n(z)$, we use the Frobenius method. After substituting Eq.~(\ref{HDE-1-extra2}) into Eq.~(\ref{HDE-1-extra1}) and using this method, we obtain
\begin{equation}\label{HDE-1-extra3}
	\alpha_{c_t}\sum_{n=0}^{\infty}T_n(z)\left(\rho^{-1}\p_{\rho}+\p^2_{\rho}\right)\rho^{n+k}e^{-\alpha_\sigma\rho^2} + \alpha_{c_z}\sum_{n=0}^{\infty}T''_n(z)\rho^{n+k}e^{-\alpha_\sigma\rho^2} = 0,
\end{equation}	
where $T''_n(z)$ is the second-order derivative of $T_n(z)$ with respect to $z$. 
After calculating the derivatives in Eq.~(\ref{HDE-1-extra3}), solving
the yielded equation for $k$, which results in $k=0$, and
using Eq.~(\ref{HDE-1-extra2}) within the source term
$\mathcal{T}^{a,c, st}$ given by
Eq.~(\ref{HDE-1-extra0}), we obtain the following equation:
\begin{align}\label{HDE-2}
	T''_n(z) + [\lambda_Q - (n+1)\lambda_0]&\,T_n(z) + (n+2)^2\lambda_z T_{n+2}(z) + \lambda_\sigma\theta(n-2)\, T_{n-2}(z) = \mathcal{T}^{a,c, st}, \nonumber \\
	&\mathcal{T}^{a,c, st} = \lambda_\rho\int_{z_i}^{z_f} M_n(z,z')\,T_n(z')dz',  \quad n=0,1,...,  
\end{align}
where $\theta (n-2) = 1$ if $n \ge 2$, otherwise $\theta (n-2) = 0$, and $M_n(z, z')$ is given by
\begin{equation}\label{M_n}
	M_n(|z - z'|) = \sum_{\alpha,\beta,\gamma, \omega_0} F^{a,c,st}_{\alpha\beta,\gamma} \int_{0}^{\infty} d\rho'\rho'^{n+1}e^{-\alpha_\sigma \rho'^{2}}|I^\gamma_{\nu}(\rho',|z-z'|)|^2.
\end{equation}
The parameters in Eq.~(\ref{HDE-2}) read
\begin{align}\label{param}
	&\lambda_\rho = k_B/\alpha_{c_z}, \nonumber \\ 
	&\lambda_Q = \alpha_Q/\alpha_{c_z}, \nonumber \\
	&\lambda_z = \alpha_{c_t}/\alpha_{c_z}, \nonumber \\
	&\lambda_0 = 4\alpha_\sigma\lambda_z, \nonumber \\
	&\lambda_\sigma = \alpha_\sigma\lambda_0.
\end{align}
With the closed-form expression of $ |I^{\gamma}_{\nu} (\rho',
|z - z'|)| ^2 $ (cf. Appendix D and Eq.~(\ref{uI2})), one can integrate
over $\rho'$ and obtain the closed-form expression of $M_n (|z - z'|)$ (cf. Eq.~(\ref{M_nnn})).

In this article we develop two mutually complementary methods of solving a
system of equations defined by Eq.~(\ref{HDE-2}). Note that each of
these equations is a \textit{second-order Fredholm
  integro-differential equation}. In addition, in these equations, the unknown
temperature profile functions $T_n(z)$ are subject to boundary
conditions. For example, Dirichlet type conditions at the material
boundaries $z=z_i$ and $z=z_f$ expressed separately for
each $\rho^n$-dependent component can be written as $T_n(z_i) = T^n_0$ and
$T_n(z_f) = T^n_N$, where $T^n_{0,N}$ are a set of given values.

Analysis of the kernel $M_n(|z - z'|)$ of Eq.~(\ref{HDE-2}) shows that it
is singular \footnote{~The reason for singular behavior is discussed in Section~\ref{numerical}.} at $z = z'$, therefore, the dominant contribution to the
term $\mathcal{T}^{a,c, st}$ on the right-hand side of
Eq.~(\ref{HDE-2}) comes from a narrow region surrounding the
singularity. If the characteristic spatial scale of temperature
variations is much larger than the characteristic radius of the
singularity region, then
\begin{equation}
  \int_{z_i}^{z_f} M_n(|z-z'|)\,T_n(z')\,dz' \approx
  T_n(z)\int_{z_i}^{z_f} M_n(|z-z'|)\,dz'.
\end{equation}
This allows us to reduce the system of integro-differential equations
to a system of ordinary differential equations (ODE). In
Section~\ref{numerical} we solve such ODE system for a particular
radiative-conductive heat transfer problem.

On the other hand, when such a simplification is not applicable, to proceed
with the solution we employ the \textit{quadrature-difference method}
that exploits a combination of composite Simpson’s one third rule and
the second-order finite differences. In order to approximate the
second-order derivatives of $T_n(z)$ with respect to $z$ based on the
\textit{quadrature-difference method}, we rewrite the left-hand side
of Eq.~(\ref{HDE-2}) in terms of forward, central, and backward
differences. After some straight-forward algebraic and matrix
operations (see Appendix G), the final thermal diffusion matrix
equation reads

\begin{align}\label{HDE-Mat}
	&\ubO\,^n \cdot \vbT^n + P^{n+2} \vbT^{n+2} + Q^{n-2} \vbT^{n-2} = \vbU^n, \qquad n = 0,1,2,...,& \nonumber \\
	&P^{n+2} = 12(n+2)^2\lambda_z h^2, \qquad Q^{n-2} = 12 \theta(n-2)\lambda_z h^2,
\end{align}
where $h = (z_i - z_f)/N$ is the step size. The $(N-1) \times (N-1)$
matrix $\ubO\,^n$ and $(N-1)$-dimensional column vectors $\vbT^n$ and
$\vbU^n$ have been introduced in Appendix G.  It has to be noted that
in some cases (the study of which is out of the scope of this work)
the system of Eqs.~(\ref{HDE-Mat}) may become ill-conditioned,
which may lead to unstable solutions demonstrating unphysical
behavior. If such a case is uncovered in practice, it is advised first to
obtain a qualitative solution with the method outlined in
Section~\ref{numerical}, and then, based on the obtained results, try to
adjust the parameters that control numerical accuracy of the
finite-difference approach (i.e., the discretization step, the rank of the
system, etc.).

Thus, with either of the methods, one can solve the
(integro-)differential equations for each order of temperature,
$T^n(z)$, and then use Eq.~(\ref{HDE-1-extra2}) to obtain an
expression for the temperature profile. With the temperature profile
at hand, one can find the radiative heat flux, Eq.~(\ref{fqrad}), and,
in turn, the radiative heat conductivity as well.  In the next
section, we present numerical results for the temperature profile in a
typical nanolayer for a few orders of $n$.

\section {\label{numerical}Numerical results and discussion}

To simulate the temperature profile, we assume DDANM is formed by
silica (SiO$_2$) and germanium (Ge) nanolayers with volume fraction of
Ge layers $f = 0.5$. Without any loss of generality, here we consider only the contribution of the $p$-polarized (i.e., $\gamma = p$) electric-electric component (i.e., $\alpha\beta = ee$) of the EDGF to the pararaxial RHT in such
uniaxial media. Eq.~(\ref{HDE-Mat}) is solved for three orders of our
temperature expansion, Eq.~(\ref{HDE-1-extra2}), that is, $n=0,1,$ and
$2$. We choose the International System of Units, SI, to express values of
parameters and quantities in our simulation, however, we scale them by
setting 1~$\mu$m as the base unit of length and 1~$\mu$s as the base
unit of time. For example, the numerical values of vacuum permittivity
and permeability in such a scaled SI system are
$\vep_0 \approx 8.854\times10^{-6}$ and
$\mu_0 \approx 1.257\times10^{-12}$, respectively. Additionally, in
this system, 1~eV~$= 1.602\times 10^{-19}$. The numerical values of other parameters
are set as follows:
\begin{align}
	&T_{00} = 700, \quad \omega_0 = 2\pi c_0/\lambda_{max}, \quad \Delta\omega = \omega_0/10, \nonumber\\ 
	&\mbox{SiO}_2: \, \,
	\alpha_c = 1.38\times 10^{-6}, \quad \vep_r = 3.90, \quad \omega_p = 0, \quad \Gamma = 0, \nonumber \\
	&\mbox{Ge}: \,\,
	\alpha_c = 64\times10^{-6}, \quad \vep_r = 16, \quad \omega_p =
   10 \, {\rm eV}/\hbar, \quad \Gamma = 1\, {\rm eV}/(2\pi\hbar),	
\end{align}
where $c_0 = 3\times 10^8$, $\vep_r$, and $\hbar = 1.055\times 10^{-28}$ are the
speed of light in vacuum, the relative permittivity, and the Planck constant in the scaled SI system, respectively, $\lambda_{max}=b/T$ is the characteristic
thermal wavelength with $b = 2.898\times 10^3$ being Wien's constant
(in the same scaled SI units), and $\omega_p$ and $\Gamma$ are the
plasma frequency and the collision frequency in the Drude dispersion
model. In the following calculations, we set the Gaussian profile
width parameter to $\sigma_{\rho} = 1$.

 \begin{figure}[hbt!]
	\centering
	\includegraphics[scale=.45]{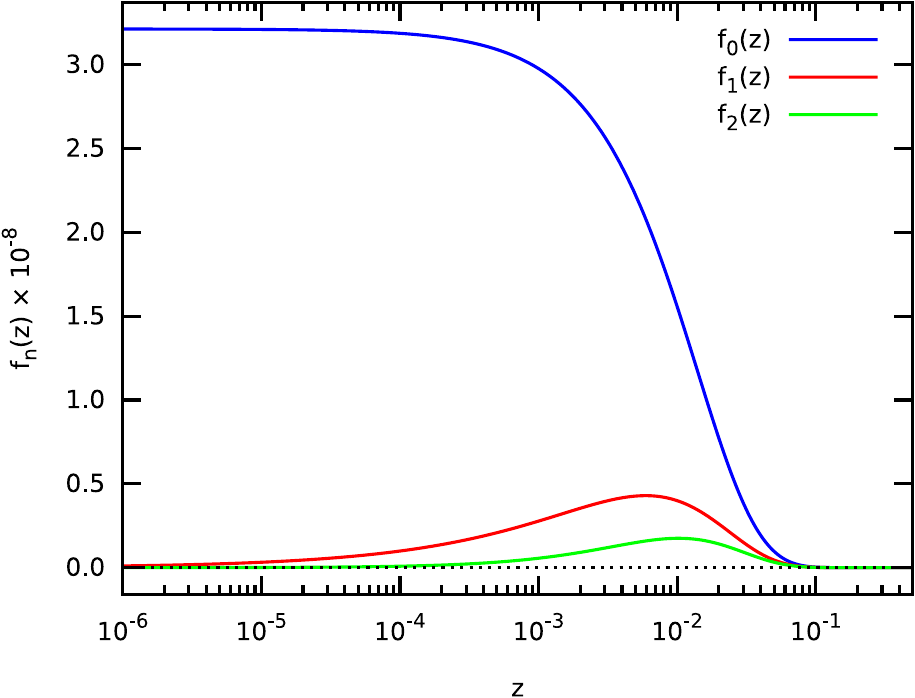}
	\caption{ The behavior of functions $f_n(z) = z M^p_n(z)$ at $z > 0$ for $n = 0, 1, 2$. (see text).}
	\label{4thfig}
\end{figure}

As was already noted, $M_n(|z-z'|)$ has a singularity at $z' = z$ [see
Eq.~(\ref{M_nnn})]. This singularity is related to the Green's
function singularity at the source location. In the context of RHT, it
can be shown that the convolution-type integrals for the radiative
heat flux and related quantities, which are expressed through such
dyadic Green functions, can be made convergent in the source
region if the Dirac-delta spatial correlation function
$\delta(\vec{r}\,'-\vec{r}\,'')$ [see Eq.~(\ref{115})] is replaced by
a correlation function with arbitrarily small, but non-zero
correlation radius (see, e.g., \cite{60}).

In the context of present work, convergence issues related to the
mentioned singularity arise only in the integrals depending on
$M_0(|z-z'|)$. Integrals involving $M_n(|z-z'|)$ with $n > 0$ all have
integrable singularities at $z = z'$ and therefore do not require any
special treatment. The singular integrals involving $M_0(|z-z'|)$ can
be evaluated by introducing a healing function equivalent to limiting
the correlation radius of the thermally fluctuating sources.  Such a
modification makes the related integrals converging in the
source region. We reserve discussing details of such a regularization
method for a future work. In the present work, we instead apply an
{\it ad-hoc} regularization of the mentioned integrals, which is based
on truncation of the non-convergent terms in a vicinity of the
singularity when $|z-z'| < \epsilon$, where $\epsilon$ is a
regularization parameter with the magnitude determined by the source
correlation radius. Indeed, Fig.~(\ref{4thfig}) shows the behavior of
functions $f_n(z) = z M^p_n(z)$ at $z > 0$ for $n = 0, 1, 2$. Here,
$M^p_n(z)$ is given by
\begin{equation}\label{MP0}
	M^p_n(z) = 4\pi^2 e^{-2\kappa^{''}_a z} F^{a,c,st}_{ee,p} \sum_{i=0}^{2}\mathfrak{M}_{i,n}^p (z),
\end{equation}
which, according to Eq.~(\ref{M_nnn}), is due to the $p$-polarized
electric-electric component of the paraxial EDGF. The quantities
$\mathfrak{M}^p_{0,n}, \mathfrak{M}^p_{1,n},$ and
$\mathfrak{M}^p_{2,n}$ are given by Eq.~(\ref{M_n012-1-SM}).
As can be seen in Fig.~(\ref{4thfig}), $M^p_0(z)$ has a $1/z$ behavior
when $z \rightarrow 0$ and therefore the corresponding integral requires
reqularization, while the other two functions tend to zero when
$z\rightarrow0$, thus requiring no regularization.

Here we will make a simplification in
solving the system of integro-differential Eqs.~(\ref{HDE-2}), in
order to reduce it to a system of coupled second-order ODEs, as
explained in the previous section. Since such a reduced system admits
a semi-analytical solution, it eliminates any potential instabilites
associated with the finite difference-based methods.

Considering the numerical parameters defined above, the reduced ODE
system can be written in the following matrix form:
\begin{equation}
  {d^2\over dz^2}
  \begin{pmatrix}
    T_0(z)\\T_1(z)\\T_2(z) \end{pmatrix} =
  \begin{pmatrix}
      a_{11} &  0  & a_{13}\\
      0     &a_{22}& 0\\
      a_{31}&   0  & a_{33}
    \end{pmatrix}\cdot
  \begin{pmatrix}
    T_0(z)\\T_1(z)\\T_2(z) \end{pmatrix},
\end{equation}
where the numerical values of the elements of the system matrix
${\bf A}$ are $a_{11} \approx 24.20$, $a_{13} \approx -48.40$,
$a_{22} \approx 48.40$, $a_{31} \approx -12.10$, and
$a_{33} \approx 72.60$, which can be obtained by evaluating the
factors at the corresponding terms of Eq.~(\ref{HDE-2}). The general
solution of the above ODE system can be written as
\begin{equation}
  \vec{\bf T}(z) = e^{-z\sqrt{\bf A}}\cdot\vec{\bf C}_1 +
  e^{+z\sqrt{\bf A}}\cdot\vec{\bf C}_2,
\end{equation}
where $\vec{\bf C}_{1,2}$ are arbitrary constant vectors, which can be
determined by taking into account the boundary conditions. In the
following calculations, we assume an externaly imposed temperature
profile at $z = 0$ and consider the radiative-conductive heat transfer
into the half-space $z > 0$. In this case, $\vec{\bf C}_2 = 0$,
and thus the temperature profile functions $T_n(z)$ encode the difference
between the local temperature at a given cross-section and the
equilibrium (e.g., ambient) temperature at $z \rightarrow +\infty$.

\begin{figure}[hbt!]
	\centering
	\subfigure[]{\includegraphics[width=.44\textwidth]{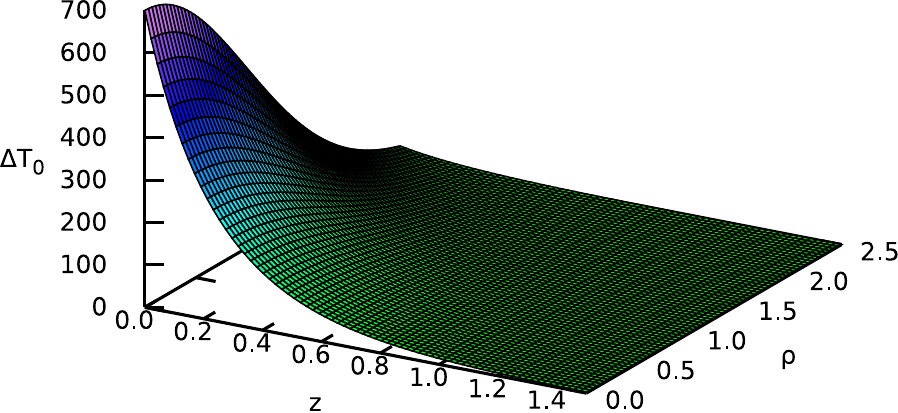}}
	\subfigure[]{\includegraphics[width=.44\textwidth]{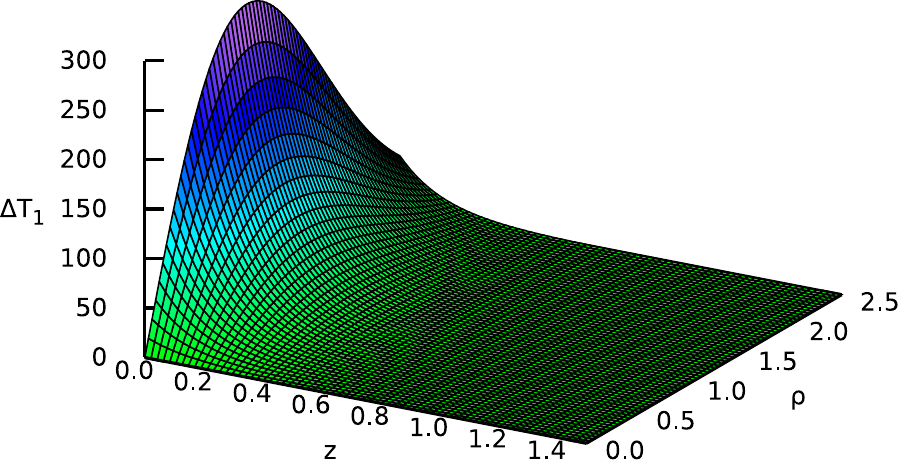}}
	\subfigure[]{\includegraphics[width=.44\textwidth]{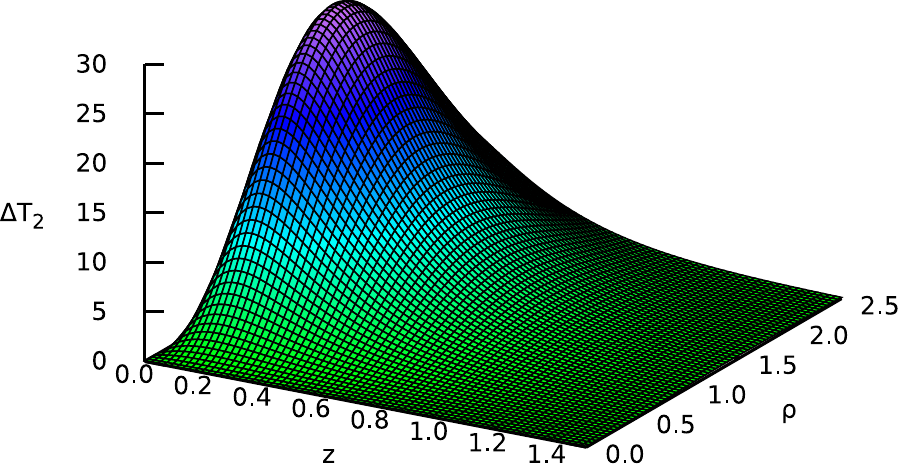}}
	\subfigure[]{\includegraphics[width=.44\textwidth]{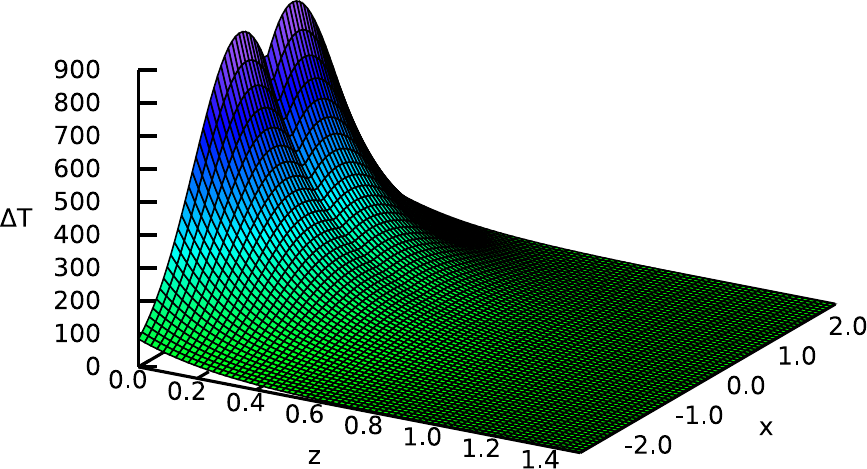}}
	\caption{ $(a)\, \Delta T_0(\rho, z)$, $(b) \, \Delta T_1(\rho, z)$, $(c)\, \Delta T_2(\rho, z)$, and $(d)\, \Delta T(x, z)$.}
	\label{5thfig}
\end{figure}

We use the Sylvester's formula to evaluate functions of martices. The
results for $\Delta T_n$, which are the separate addends in expanded
Eq.~(\ref{HDE-1-extra2}) that correspond to terms with $n=0,1,2$,
respectively, are depicted in Figs.~(\ref{5thfig})-(a,b,c). In these calculations, we set
$\vec{\bf C}_1=(T_{00},0.7\times T_{00}, 0)^T$, which defines the
imposed temperature profile at $z = 0$. As one can see, the $n = 0$
term of the temperature expansion makes the
dominant contribution to the temperature profile. The $n = 1$ term is
decoupled from the other two terms and decays away from the interface
$z = 0$. On the other hand, the $n = 2$ term, which is zero at
$z = 0$, initially grows with $z$ due to coupling to the $n=0$ term,
reaches a maximum and then decays. All terms decay at
$z \rightarrow +\infty$. Fig.~(\ref{5thfig})-(d) shows the total
temperature profile in the $xz$-plane, which is a superposition of the
three profiles.  As one can see, the total profile when plotted in the
$xz$ plane, has a distinctive feature at $x = 0$, which is due to the
presence of the $n=1$ term. Although in these calculations we have ignored the orders
higher than $n=2$, the numerical
results demonstrate expected physical behavior, which confirms physical
soundness of the developed theoretical framework.

\section{Conclusion}

By employing the novel concept of EDGF, rather than the standard DGF
technique, we have developed a self-consistent theoretical method for
modeling the DHT in DDANM, such as MMs. To model the propagation of
thermally-agitated fluctuations of electromagnetic fields through
DDANM, which involves modeling of dynamic heat storage and release
due to both photonic and phononic processes, we assumed that the
photonic radiative heat transfer mechanisms in DDANM are
complemented with dynamic phononic mechanisms of heat storage and
conduction, taking into account the effects of local heating, heat
storage, and release. Furthermore, we have
developed a theoretical framework for studying the propagation of
quasi-monochromatic signals through dispersive and dissipative
media. By representing the time-dependent EM fields associated with
such processes as products of the slowly varying amplitude and the
quickly oscillating carrier, we have formulated a system of equations
for the SVAs of the electromagnetic fields in such media.

We solved the macroscopic Maxwell equations using this method and
employed the Poynting theorem and fluctuation-dissipation theorem to
explain the excitation and propagation of FEFs. We obtained a closed
form expression for the coupling term of the photonic-phononic
mechanism by using macroscopic electro- and thermodynamics. Then, we
applied the yielded closed-form relations for modeling of the paraxial
radiative heat transfer in DDAN uniaxial MMs. By considering a
Gaussian form for the transverse temperature profile, we obtained a
second-order Fredholm integro-differential equation for heat diffusion
in the steady-state case, which can be used to model the paraxial
radiative-conductive heat transfer in such media. Using the
quadrature-difference method, we solved this integro-differential
equation. In addition, we have also developed a complimetary
analytical method to solve for the termal profile in the case when the
Fredholm equation reduces to an ODE. When the temperature profile is
known, one can calculate the radiative and conductuve heat fluxes and,
in turn, the effective heat conductivity due to both mechanisms.

In the numerical examples we have considered application of the
developed analytical method to radiative-conductive heat transport in
nanolayered media formed by layers of Ge and SiO$_2$. We have obtained
the temperature profiles for the three first orders of expansion and
the total temperature profile, for a scenario when external heat penetrates
into the halfspace $z > 0$ occupied by such unuaxial material. We have
also discussed singularities resulting from the idealized assumption
that the thermal fluctuations at arbitrarily close points in space are
fully uncorrelated.

With minimal adaptations, the developed framework can be used to model
coupled electromagnetic wave and heat propagation through a material
whose parameters may depend on the local temperature, which is
affected by the passing electromagnetic radiation. We plan to present
such an extension of this method in a future work.

The results of this research may find applications in the areas of
science and technology related to thermophotovoltaics, energy
harvesting, RHT cooling, and microfluidic/nanofluidic thermal
management, and others.

\section*{Acknowledgments}
The authors acknowledge financial support by FCT---Funda\c{c}\~{a}o
para a Ci\^{e}ncia e Tecnologia, I.P., by project Ref.~UIDB/50008/2020
and DOI identifier 10.54499/UIDB/50008/2020,
https://doi.org/10.54499/UIDB/50008/2020, and by the previous project
Ref.~UID/EEA/50008/2013, sub-project SPT.

\section*{Appendix A: closed-form of $Q^\alpha_{EM}$}

The total heat produced by the dissipative processes per unit volume of the material, has electric and magnetic parts: $Q_{EM} \equiv \sum_{\alpha = e, m} Q^{\alpha}_{EM}$, (see Eq.~(\ref{106})). First, let us find a closed-form for $Q^{\,e}_{EM}$. Using Eq.~(\ref{15}), we can write the above equation as $Q^{\,e}_{EM} \equiv Q^{\,e}_{EM,\,1} + Q^{\,e}_{EM,\,2}$, where
\begin{equation}\label{106-extra}
	\begin{aligned}
		&Q^{\,e}_{EM,\,1} = {1 \over 2} \Re\! \int\limits_{-\infty}^{\infty}\p_{t} \left[\left(\uze + \p_{\omega}\uze i\p_{t}\right)_{_{\omega_0}} \cdot\vF_m^e \right]^\dagger\cdot\vF_m^e\,dt  \\ 
		&Q^{\,e}_{EM,\,2} = {1 \over 2} \Re\! \int\limits_{-\infty}^{\infty}(i\omega_0)\left[\left(\uze + \p_{\omega}\uze i\p_{t}\right)_{_{\omega_0}} \cdot\vF_m^e \right]^\dagger\cdot\vF_m^e\,dt	 
	\end{aligned}
\end{equation}
By ignoring the terms on the order of $O(\p^2_{t})$ and higher and separating the
material dyadic operator $\uze=\uzep + i\uzepp$ into the Hermitian and
anti-Hermitian parts, $\uzep$ and $i\uzepp$, respectively, one can
easily verify that
$\Re \int_{-\infty}^{\infty}
\vF^{e\,\dagger}_{m}\cdot(\uzep\p_{t})\cdot\vF_{m}^{e}\,dt = 0$ since
$\p^\dagger_{t}=-\p_{t}$ and then
\begin{equation}\label{106-extra-extra-1}
	Q^{\,e}_{EM,\,1} = {1 \over 2} \int\limits_{-\infty}^{\infty}\vF_m^{\,e\dagger}\cdot(\uzepp_{_{\omega_0}}i\p_{t})\cdot\vF_m^e\,dt.
\end{equation}
In the above equation, at the last step, we used the fact that
$ \int\limits_{-\infty}^{\infty}i\p_{t}
\left(\vF_m^{\,e\dagger}\cdot\uzepp_{_{\omega_0}}\cdot\vF_m^e\right)\,dt= 0$.  Now, we can decompose $Q^e_{EM,\,2}$ in two terms as follows
\begin{equation}\label{106-extra-extra-2}
	\begin{aligned}
		Q^{\,e}_{EM,\,2} =\,
		&{1 \over 2} \Re\int\limits_{-\infty}^{\infty}(i\omega_0)\left[ (\uzep+i\uzepp)_{_{\omega_0}}\cdot\vF_m^{\,e}\right]^\dagger \cdot\vF_m^{\,e}\,dt \\
		&+\,{1 \over 2} \Re\int\limits_{-\infty}^{\infty}(i\omega_0)\left[ \p_{\omega}(\uzep+i\uzepp)_{_{\omega_0}}\cdot (i\p_{t})\vF_m^{\,e}\right]^\dagger\cdot\vF_m^{\,e}\,dt,
	\end{aligned}
\end{equation}
where the first term results in
${1 \over 2}\int\limits_{-\infty}^{\infty} \vF_m^{\,e\dagger}
\cdot(\omega\uzepp)_{_{\omega_0}}\cdot\vF_m^{\,e}\,dt $ and by
considering $i\p_{t}$ as a Hermitian operator, the second term leads
to
${1 \over 2} \int_{-\infty}^{\infty}\vF_m^{e\,\dagger}\cdot
(\omega\p_{\omega}\uzepp)_{_{\omega_0}}\cdot(i\p_{t}\vF_m^{e})\,dt$.
Thus, summing all the above results, Eq.~(\ref{106-extra}) becomes
\begin{equation}\label{107-extra}
	Q^{\,e}_{EM} = {1 \over 2} \int\limits_{-\infty}^{\infty} \vF_m^{\,e\,\dagger}\cdot\left[ \omega\uzepp + \p_{\omega}(\omega\uzepp) (i\p_{t})\right]_{_{\omega_0}}\cdot\vF_m^{\,e}\,dt.
\end{equation}
A similar result can be obtained for the $Q^{\,m}_{EM}$ in
Eq.~(\ref{106}). Therefore, we can write
\begin{equation}\label{108-A}
	Q^\alpha_{EM} = {1 \over 2} \int\limits_{-\infty}^{\infty}
	\vF_m^{\,\alpha\dagger}\cdot\left[ \omega\uzpp\,^\alpha +
	\p_{\omega}(\omega\uzpp\,^\alpha)
	(i\p_{t})\right]_{_{\omega_0}}\cdot\vF_m^{\,\alpha}\,dt,
	\qquad (\mbox{where}\ \alpha\ \mbox{is}\ e\ \mbox{or}\ m)
\end{equation} 

\section*{Appendix B: Extracting PEDGF}

To give an extract of the the EDGF formalism, we consider uniaxial
media with the material dyadics of the form
\begin{equation}\label{37}
	\uza=\zt^\alpha\uI_{t}+\zz^\alpha\uI_{z},  \qquad 	\alpha = e \ \mbox{or} \ m, 
\end{equation}
with $\uI_{t}=\hat{x}\hat{x}+\hat{y}\hat{y}$ and $\uI_{z}=\hat{z}\hat{z}$ as the identity dyadics in two orthogonal directions: the tangential and axial, where we can decompose the propagation wave vector as $\vk\times\uI=(\vk_{t}+\vk_{z})\times(\uI_{t}+\uI_{z})$.
From the EDGF formalism of~\cite{60}, the envelope dyadic Green's function in the cylindrical coordinates, as sum of two orthogonal polarizations, TE($p$) and TM($s$), is given by
\begin{align}\label{EDGF}
	\ubG(k_{t},\vvro,\pm|Z|;\Dom,\tau) = { \Dom \over (2\pi)^{3} } \sum_{\gamma=p,s}&\bigg\lbrace \int d\vk_{t} e^{i(\vk_{t}\cdot\vvro + \kappa^{\gamma}_{0}|Z|)} \nonumber \\
	&\times \left[ j_{0}(\tau_{g}^{\gamma}\Dom/2)\, \ubC - 
	i{\Dom \over 2} j_{1}(\tau_{g}^{\gamma}\Dom/2)\, \ubD \right] \bigg\rbrace_{\omega_0},
\end{align}
where $\vvro = \vro - \vro^{\,\prime}$ and
$|Z|=|z-z^{\,\prime}|$. In the above equation, we have
expanded $\vec{r}$ and $\vec{r}^{\,\prime}$ as
$\vec{r}=(\vro,\vec{z}\,)$ and
$\vec{r}^{\,\prime}=(\vro^{\,\prime},\vec{z}^{\,\prime})$,
respectively. The notations $j_{0}(\cdot)$ and $j_{1}(\cdot)$ represent the zeroth- and first-order spherical Bessel functions, and $\tau_{g}^{\gamma} = \tau-Z/V^{\gamma}_{g}$ with $V^{\gamma}_{g}$ as the group velocity of FEFSVA. The propagation factor $\kappa^{\gamma}_{0}$ and the group velocity read
\begin{align}
	\kappa^{p[s]}_{0}=\sqrt{a^e_{t}a^m_{t}-a^{e[m]}_{t}\kt2/a^{e[m]}_{z}}, \quad \Im(\kappa_{p[s]})>0, \label{66} \\ 
	V^{p[s]}_{g}={2\kappa^\gamma_{0} \over (a^{e}_{t}b^{m}_{t}+a^{m}_{t}b^{e}_{t})-(b^{e[m]}_{t}a^{e[m]}_{z}-a^{e[m]}_{t}b^{e[m]}_{z})\kt2/a^{e[m]^2}_{z}}, \quad  \label{67}
\end{align}
where  
\begin{align}
	a^e_{i}=\omega\vep_{i},  \qquad b^e_{i}=\p_{\omega}(\omega\vep_{i}), \qquad \mbox{(TM)}, \label{56} \\
	a^m_{i}=\omega\mu_{i},   \qquad b^m_{i}=\p_{\omega}(\omega\mu_{i} ), \qquad \mbox{(TE)}, \label{57}
\end{align}
with the index $i$ denotes $t$ and $z$. 
Reader can find the components of $\ubC$ and $\ubD$ in~\cite{60}.

Now, for the paraxial propagation case, the propagation factor and the group velocity are given in~\cite{60}:
\begin{align}\label{132}
	\kappa^{p[s]}_{0} = \sqrt{\kappa_a^2-{a^{e[m]}_{t}\kt2 \over a^{e[m]}_{z}}}, \qquad
	V_g^a = {2\kappa_a \over (a_{t}^{e}b^{m}_{t}+a^{m}_{t}b^{e}_{t})},
\end{align}
where $\kappa_{a}=\pm \sqrt{a^e_{t}a^m_{t}}$, with $\Im(\kappa_a)>0$, in which each of $+$ or $-$ should be selected such that $\Im(\kappa_a)>0$. By keeping this point in mind, we set $\kappa_{a}=+\sqrt{a^e_{t}a^m_{t}}$ in our notations. 
As can be seen in Eq.~(\ref{132}), $V^a_g$ is no longer dependent on the polarization of the propagation and also $k_t$. Thus, we can find $\tau^\gamma_{g, paraxial}\equiv\tau_{a}=\tau-|Z|/V^a_g$ is also independent of polarization and $k_t$. 
When $\left|a^{e[m]}_{z}\kt2\right|\ll \left|a^{e[m]}_{z}\kappa_a^2\right|$, we can expand $\kappa_0^{p[s]}$
as follows
\begin{equation}\label{133}
	\kappa^{p[s]}_{0} \approx \pm \bigg(\kappa_{a}-{1\over 2}{a^{e[m]}_{t}\kt2\over a^{e[m]}_{z}\kappa_{a}}\bigg), \qquad \Im(\kappa^{p[s]}_0) > 0.
\end{equation}
Regarding independence of $V^a_g$ and $\tau_{a}$ on $k_t$ and
polarization, and setting $k_t \approx 0$ in $\ubC$ and $\ubD$
\cite{60}, the PEDGF,
$\ubG\,^a(k_{t},\vvro,\pm|Z|;\Dom,\tau)$, can be reduced to
\begin{equation}\label{PEDGF-0}
	\begin{aligned}
		\ubG\,^a = { \Dom \over (2\pi)^{3} } \bigg\lbrace  &e^{i\kappa_a|Z|} \sum_{\gamma=p,s} \int d\vk_{t} e^{i(\vk_{t}\cdot\vvro - \chi^{\gamma}_{a}k_t^2)}  \\
		&\times \left[j_{0}(\tau_{a}\Dom/2)\, \ubC\,^{\gamma}_a  - 
		i{\Dom \over 2}j_{1}(\tau_{a}\Dom/2)\, \ubD\,^{\gamma}_a \right] \bigg\rbrace_{\omega_0},
	\end{aligned}
\end{equation}
where $\gamma = p$ or $s$, $\chi^{p[s]}_{a}={a^{e[m]}_t\over 2a^{e[m]}_z\kappa_a}|Z|$, and $\ubC\,^\gamma_a$ and $\ubD\,^\gamma_a$ are given by:
\begin{equation}\label{135}
	\ubC\,^{\gamma}_a = 
	\begin{bmatrix}
		\uC\,^{tt,\gamma}_a && \u0 \\
		\u0 && \u0
	\end{bmatrix}, \qquad
	\ubD\,^{\gamma}_a = 
	\begin{bmatrix}
		\uD\,^{tt,\gamma}_a && 0 \\
		\u0 && \u0
	\end{bmatrix}.
\end{equation}
Removing superscript $tt$ from the different matrix elements
of $\uC\,^{tt}_a$ and $\uD\,^{\,tt}_a$, the dyadic matrix
forms, denoted here by $\uAa_a$ expressed as
\begin{equation} \label{136}
	\uAa_a\equiv
	\begin{bmatrix}
		\mathfrak{A}^{a,p}_{ee}\, \ue\,^{p}_{ee} && \mathfrak{A}^{a,p}_{em}\, \ue\,^{p}_{em}\\
		\mathfrak{A}^{a,p}_{me}\, \ue\,^{p}_{me} && \mathfrak{A}^{a,p}_{me}\, \ue\,^{p}_{mm}
	\end{bmatrix} \,+ 
	\begin{bmatrix}
		\mathfrak{A}^{a,s}_{ee}\, \ue\,^{s}_{ee} && \mathfrak{A}^{a,s}_{em}\, \ue\,^{s}_{em}\\
		\mathfrak{A}^{a,s}_{em}\, \ue\,^{s}_{em} && \mathfrak{A}^{a,s}_{mm}\, \ue\,^{s}_{mm}
	\end{bmatrix}, 
\end{equation}
where the different non-zero elements of $\mathfrak{A}^{a,\gamma}_{\alpha\beta}$, with $\mathfrak{A}=C,D$, are calculated by setting $k_t\approx0$ for different elements of $\ubC$ and $\ubD$ \cite{60}, we obtain
\begin{gather}\label{pCD's}
	C^{a,\gamma}_{ee[mm]}=-{a^{m[e]}_{t} \over 2\kappa_{a}}, \qquad C^{a,\,s[p]}_{em[me]} =[-]{n_z\over 2},\\
	D^{a,\gamma}_{ee[mm]}=-[+]{a^{m[e]}_{t} \over 4\kappa_{a}}\left( {b_t^{m}\over{a_t^{m}}} - {b_t^{e}\over{a_t^{e}}}\right),
\end{gather}
with $n_{z} = +1\, or -1$, depending on the sign of $Z$, that is,
$Z>0\, or\, Z<0$. In Eq.~(\ref{136}), $\ue\,^\gamma_{\alpha\beta}$ are given by \cite{60}
\begin{equation}\label{pcdhat-0}
	\ue\,^{p[s]}_{ee[mm]}=\htt\htt, \qquad	\ue\,^{p[s]}_{mm[ee]}=\hs\hs, \qquad
	\ue\,^{p[s]}_{em}=\htt\hs, \qquad	\ue\,^{p[s]}_{me}=\hs\htt,
\end{equation}
where $\htt= \vec{k_t}/k_t$ and $\hs= \hz \times \htt$. 
Now, remembering that the PEDGF block matrix has only the transverse-transverse block element, the general form of PDGF elements is given by
\begin{align}\label{PEDGF}
	\uG\,^{a,\gamma}_{\alpha\beta} = [\Dom/(2\pi)^{3}]\, & \left[C\,^{a,\gamma}_{\alpha\beta} j_{0}(\tau_a\Dom/2) - 
	i(\Dom/2) D\,^{a,\gamma}_{\alpha\beta} j_{1}(\tau_a\Dom/2)\right]\,\uI\,^{\gamma}_{\alpha\beta}(\vec{\varrho},|Z|), \nonumber \\
	& \uI\,^{\gamma}_{\alpha\beta}(\vec{\varrho},|Z|) = e^{i\kappa_a|Z|}\int \ue\,^{\gamma}_{\alpha\beta}\, e^{i[\vk_{t}\cdot\vec{\varrho}-\chi^{\gamma}_{a}k_t^2]}\, d\vk_{t}.
\end{align}

\section*{Appendix C: closed-form of $\Tr\left[\ubGt\,^{\gamma}_{\alpha\beta}\right]_{_{\omega_0}} j_{_0}(\tau''')$}

To find a closed-form of the integrand in Eq.~(\ref{phrat}), first, we expand $\Tr[\ubGt]_{_{\omega_0}}j_{_0}(\tau''')$ by allowing
$\ubA_s(t)$ and $\ubB_{s}(t')$ to operate on $\ubG\,^{a\,\dagger}$, $\ubG\,^a$, and $j_{_0}(\tau''')$.
According to Eqs.~(\ref{Comp-PEDGF1})-(\ref{Comp-PEDGF4}), we can write
\begin{align}\label{140}
	\Tr\left[\ubGt\,^{\gamma}_{\alpha\beta}\right]_{_{\omega_0}} & j_{_0}(\tau''')= \nonumber \\
	&\Tr\left[ \uG\,^{\,a,\gamma\,\dagger}_{\alpha\beta}(t,t'')\cdot\uA\,^{\alpha}_{s,t}(t)\cdot \uG\,^{\,a,\gamma}_{\alpha\beta}(t,t') \cdot\uB\,^{\beta}_{s,t}(t')\right]_{_{\omega_0}}j_{_0}(\tau''').
\end{align}	
Recalling Eqs.~(\ref{110}) and (\ref{118}) and expanding $\uA\,^{\alpha}_{s,i}$ and $\uB\,^{\alpha}_{s,i}$, with $i=t,z$, as
\begin{equation}\label{141-1}
	\uA\,^{\alpha}_{s,i} = (a^\alpha_{0_i} + a^\alpha_{1_i}\overleftrightarrow{i\p_t})\uI_{i}, \qquad a^{e[m]}_{0_i}=\omega\vep''_{i}[\mu''_{i}], \qquad a^{e[m]}_{1_i}=\p_{\omega}(a^{e[m]}_{0_i}),
\end{equation}	
\begin{equation}\label{141-2}
	\uB\,^{\alpha}_{s,i} = (b^\alpha_{0_i} + b^\alpha_{1_i}\overleftrightarrow{i\p_{t'}})\uI_{i},\qquad
	b^{e[m]}_{0_i}=\omega\vep''_{i}[\mu''_{i}]\Theta, \qquad b^{e[m]}_{1_i}=\p_{\omega}(b^{e[m]}_{0_i}),
\end{equation}	
and working with the tangential components, $\uI_{i}=\uI_{t}$, we obtain
\begin{align}\label{142}
	\Tr[\ubGt\,^{\gamma}_{\alpha\beta}]_{_{\omega_0}}j_{_0}(\tau''')&= \nonumber \\ 
	\sum_{\alpha,\beta,\gamma} & \left[ \uG\,^{a,\gamma\,\dagger}_{\alpha\beta}(t,t'')(a^\alpha_{0_t} + a^\alpha_{1_t}\overleftrightarrow{i\p_t}):\uG\,^{a,\gamma}_{\alpha\beta}(t,t')(b^\alpha_{0_t} + b^\alpha_{1_t}\overleftrightarrow{i\p_{t'}})\right]_{_{\omega_0}}j_{_0}(\tau'''),
\end{align}
where we used the relation of $\Tr(\uA\cdot\uB)=\uA:\uB$. Now, according to Eq.~(\ref{145}), one can expand Eq.~(\ref{142}) as follows
\begin{align}\label{146-extra}
	\Tr\left[\ubGt\,^{\gamma}_{\alpha\beta}\right]_{_{\omega_0}} j_{_0}(\tau''') &= \sum_{\alpha,\beta,\gamma} \bigg[ 
	a^\alpha_{0_t}b^\alpha_{0_t} \uG\,^{a,\gamma\,\dagger}_{\alpha\beta}(t,t''):\uG\,^{a,\gamma}_{\alpha\beta}(t,t') j_{_0}(\tau''') \nonumber \\
	&\, + \, a^\alpha_{0_t}b^\alpha_{1_t} \uG\,^{a,\gamma\,\dagger}_{\alpha\beta}(t,t''):\uG\,^{a,\gamma}_{\alpha\beta}(t,t')(\overleftrightarrow{i\p_{t'}})j_{_0}(\tau''')  \nonumber \\
	&\, + a^\alpha_{1_t}b^\alpha_{0_t} \uG\,^{a,\gamma\,\dagger}_{\alpha\beta}(t,t'')(\overleftrightarrow{i\p_{t}}):\uG\,^{a,\gamma}_{\alpha\beta}(t,t') j_{_0}(\tau''')\bigg]_{_{\omega_0}},
\end{align}
where we have neglected the second-order time-derivative terms such as $\p_{t}(...)\p_{t'}(...)$. By directly computing time derivatives with respect to $t$ and $t'$, we arrive at following results:
\begin{equation}\label{146-extra-extra-1}
	i\p_{t'} j_0(\tau''') = -i{\Dom \over 2} j_1(\tau'''), \qquad i\p_t \uG\,^{a,\gamma}_{\alpha\beta}(t,t') = - i{\Dom \over 2} \uH\,^{a,\gamma}_{\alpha\beta}(t,t^{\pr}),
\end{equation}
\begin{equation}\label{146-extra-extra-2}
	i\p_{t'} \uG\,^{a,\gamma}_{\alpha\beta}(t,t') = + i{\Dom \over 2} \uH\,^{a,\gamma}_{\alpha\beta}(t,t^{\pr}), \quad
	i\p_{t} \uG\,^{a,\gamma\,\dagger}_{\alpha\beta}(t,t'') = - i{\Dom \over 2}\uH\,^{a,\gamma\,\dagger}_{\alpha\beta}(t,t^{\pr\pr}),
\end{equation}
where $j_1(x) \equiv \sin(x)/x^2 - \cos(x)/x$ is the first-order spherical Bessel function, and $\uH\,^{a,\gamma}_{\alpha\beta}(t,t^{\pr})$ are given by
\begin{equation}\label{147} 
	\uH\,^{a,\gamma}_{\alpha\beta}(t,t^{\pr})= {\Dom \over (2\pi)^3}h^{a,\gamma}_{\alpha\beta}(\tau'_{a})\uI\,^{\gamma}_{\alpha\beta}(\vec{\varrho},|Z|).
\end{equation}
In the above equation, $\uI\,^{\gamma}_{\alpha\beta}(\vec{\varrho},|Z|)$ is given by Eqs.~(\ref{Comp-PEDGF1})-(\ref{Comp-PEDGF4}) and $h^{a,\gamma}_{\alpha\beta}(\tau'_{a})$ is defined by
\begin{equation}\label{149}
	h^{a,\gamma}_{\alpha\beta}(\tau'_{a}) = C^{a,\gamma}_{\alpha\beta} j_{1}(\tau'_{a}) + 
	i{\Dom \over 2} D^{a,\gamma}_{\alpha\beta} \big[j_{1}(\tau'_{a})/\tau'_{a}-j_2(\tau'_a)\big],
\end{equation}
where $j_2(x) \equiv (3/x^3-1/x)\sin(x) - 3\cos(x)/x^2$ is the second-order spherical Bessel function of the first kind, and all other indices and variables can be found in Eqs.~(\ref{Comp-PEDGF1})-(\ref{Comp-PEDGF4}).

Now, having Eqs.~(\ref{143}), (\ref{145}), (\ref{146-extra}), (\ref{146-extra-extra-1}), and (\ref{146-extra-extra-2}) in hand, we can write
\begin{align}\label{146-B}
	&\Tr\left[\ubGt\,^{\gamma}_{\alpha\beta}\right]_{_{\omega_0}} j_{_0}(\tau''')= \nonumber \\
	& k_BT \sum_{\alpha,\beta,\gamma}\bigg[  a^\alpha_{0_t}(a^\beta_{0_t} + a^\beta_{1_t} \dot{T}' / 2T) \left(\uG\,^{a,\gamma\,\dagger}_{\alpha\beta}(t,t''):\uG\,^{a,\gamma}_{\alpha\beta}(t,t')\right)j_{_0}(\tau''') \nonumber \\
	& - {i\Dom\over 4} a^\alpha_{1_t} a^\beta_{0_t} \left(\uH\,^{a,\gamma\,\dagger}_{\alpha\beta}(t,t''):\uG\,^{a,\gamma}_{\alpha\beta}(t,t') + 
	\uG\,^{a,\gamma\,\dagger}_{\alpha\beta}(t,t''):\uH\,^{a,\gamma}_{\alpha\beta}(t,t')
	\right)j_{_0}(\tau''')  \nonumber \\
	& + {i\Dom\over 4} a^\alpha_{0_t} a^\beta_{1_t} \bigg(\uG\,^{a,\gamma\,\dagger}_{\alpha\beta}(t,t''):\uH\,^{a,\gamma}_{\alpha\beta}(t,t')j_{_0}(\tau''') 
	- \uG\,^{a,\gamma\,\dagger}_{\alpha\beta}(t,t''):\uG\,^{a,\gamma}_{\alpha\beta}(t,t')j_{_1}(\tau''')\bigg) \bigg]_{_{\omega_0}}.	\nonumber\\
\end{align}	 

\section*{Appendix D: Calculation of $|I_\nu^\gamma(\vR)|^2$}

In order to calculate the integral over $\vk_{t}$ in
$\uI^\gamma_{\alpha\beta}(\vvro,|Z|)$, we can rewrite
Eq.~(\ref{PEDGF}) by setting $\nabla_t=i\vec{k_t}$ where
$\nabla_t$ is the gradient operator in the transverse plane, which, in cylindrical coordinates, reads $\nabla_t = \p_{\varrho}\, \hat{\varrho} +  \p_{\phi}\,\hat{\phi}/\varrho $. Therefore, we can substitute following expressions in Eq.~(\ref{PEDGF}):	
\begin{align}\label{pcdhat}
	&\ue\,^{p[s]}_{ee[mm]}=-\nabla_t\nabla_t/k_t^2, \qquad	\ue\,^{p[s]}_{mm[ee]}= - \hz \times \nabla_t \hz \times \nabla_t/k_t^2, \nonumber \\
	&\ue\,^{p[s]}_{em}=-\nabla_t\hz \times \nabla_t/k_t^2, \qquad	\ue\,^{p[s]}_{me}=-\hz \times \nabla_t \nabla_t/k_t^2.
\end{align}
As can be seen in above, all dyadic operators $\ue\,^{\gamma}_{\alpha\beta}$ contain $k_t^2$. So, 
by keeping $k_t^2$ in the integral over $\vec{k_t}$ in $\uI\,^\gamma_{\alpha\beta}(\vvro,|Z|)$ and taking the rest out of the integral, we need to solve the following simple integral
\begin{equation}\label{I_abg-0}
	I\,^{\gamma}(\vvro\,) = \int e^{i\vk_{t}\cdot\vec{\varrho}}\,\left[e^{-i\chi^{\gamma}_{a}k_t^2}/k_t^2\right] d\vk_{t},
\end{equation}
where we have not included the factor of
$e^{i\kappa_a|Z|}$. In above, the bracketed expression can be written in an integral form as $\int_{-\infty}^{-i\chi^\gamma_a} d\eta e^{\eta k_t^2}$. By setting $\eta \rightarrow -i\eta$, we rewrite Eq.~(\ref{I_abg-0}) as follows:
\begin{equation}\label{I_abg-1}
	I\,^{\gamma}(\vvro\,) = -i\int_{-i\infty}^{\chi^\gamma_a} d\eta \left[\int e^{i[\vk_{t}\cdot\vec{\varrho}-\eta k_t^2]} d\vk_{t}\right].
\end{equation}
By expanding $\vvro = \vec{X}+\vec{Y}$ and decomposing the integral over $\vk_{t}$ into two integrals over $k_x$ and $k_y$, we have
\begin{equation}\label{I_abg-2}
	I\,^\gamma(X,Y) = -i\int_{-i\infty}^{\chi^\gamma_a} d\eta \left[\int_{-\infty}^{\infty} dk_x\,e^{i(k_x\,X-\eta k_x^2)}\right] \left[ \int_{-\infty}^{\infty} dk_y\,e^{i(k_y\,Y-\eta k_y^2)}\right].
\end{equation}
Since two integrals in brackets are the same except for the dummy variables $k_x $ and $k_y$, we go on by considering one of them and have
\begin{equation}\label{I_abg-3}
	\int_{-\infty}^{\infty} dk_x\,e^{i(k_x\,X-\eta k_x^2)} = e^{iX^2/4\eta}\int_{-\infty}^{\infty} dK_x\,e^{-i\eta K_x^2}, \qquad K_x = k_x-X/2\eta.
\end{equation}
Substituting this result into Eq.~(\ref{I_abg-1}), recalling $\int_{0}^{\infty} e^{-iax^2}xdx=1/(2ia)$ with $a$ as a complex number, after some manipulation, we have
\begin{equation}\label{I_abg}
	I\,^\gamma(\varrho) = -\pi \int_{-i\infty}^{\chi^\gamma_a} d\eta e^{i\varrho^2/4\eta}/\eta.
\end{equation}
Now, we are ready to include the dyadic parts of
$\uI\,^\gamma_{\alpha\beta}(\vvro)$. Noticing that $\nabla_t$
and $\hz\times \nabla_t$ depend on $\hat{\rho}$ and
$\hat{\phi}$ and the integrand in Eq. (\ref{I_abg}) does not
depend on the azimuthal angle $\phi$, Eq.~(\ref{PEDGF}) for all components of
$\uI\,^\gamma_{\alpha\beta}(\vvro)$ except
$\uI\,^\gamma_{me}(\vvro)$ takes on the following form:
\begin{equation}\label{uI_abg-0}
	\uI\,^\gamma_{\alpha\beta} (\vvro\,) = -\pi\uu\,^\gamma_{\alpha\beta} \p_{\varrho}\p_{\varrho} \int_{-i\infty}^{\chi^\gamma_a} d\eta e^{i\varrho^2/4\eta}/\eta,
\end{equation}
where the dyadics $\uu\,^\gamma_{\alpha\beta}$ are as follows:
\begin{equation}\label{youhat}
	\uu\,^{p[s]}_{ee[mm]}=-\hat{\rho}\hat{\rho}, \qquad	\uu\,^{p[s]}_{mm[ee]}= - \hat{\phi}\hat{\phi}, \qquad
	\uu\,^{p[s]}_{em}=-\hat{\rho}\hat{\phi}.
\end{equation}
For the case of $\uI\,^\gamma_{me}(\vvro\,)$, we have
\begin{equation}\label{uI_abg-0-me}
	\uI\,^\gamma_{me} (\vvro\,) = -\pi\left(\hat{\phi}\hat{\rho}\, \p_{\varrho}\p_{\varrho} - {\hat{\rho}\hat{\phi} \over \varrho} \p_{\varrho} \right) \int_{-i\infty}^{\chi^\gamma_a} d\eta e^{i\varrho^2/4\eta}/\eta.
\end{equation}
After performing differentiation with respect to $\varrho$ and
integrating over $\eta$ in Eq.~(\ref{uI_abg-0}) while including
$|Z|$ dependency, the final form of
$\uI\,^\gamma_{\alpha\beta} (\vvro, |Z|)$ is given by
\begin{equation}\label{uI}
	\uI\,^\gamma_{\alpha\beta} (\vvro, |Z|) = 2\pi e^{i\kappa_a|Z|}\left[ \left({1 \over \varrho^2} - {i \over 2\chi_a^\gamma} \right) e^{i\varrho^2  / 4\chi_a^\gamma} - {1 \over \varrho^2} \right] \uu\,^\gamma_{\alpha\beta},
\end{equation}
and that of $\uI\,^\gamma_{me} (\vvro, |Z|)$ reads
\begin{align}\label{uI-me}
	\uI\,^\gamma_{me} (\vvro, |Z|) &= 2\pi e^{i\kappa_a|Z|} \nonumber \\ 
	&\times \left\lbrace \left[ \left({1 \over \varrho^2}
	- {i \over 2\chi_a^\gamma} \right) e^{i\varrho^2  /
		4\chi_a^\gamma} - {1 \over \varrho^2} \right]
	(-\hat{\phi}\hat{\rho}) + {1 \over \varrho^2}\left( 1 -
	e^{i\varrho^2  / 4\chi_a^\gamma} \right)
	\hat{\rho}\hat{\phi} \right\rbrace .
\end{align}

The above expressions will lead to a complicated form for
$\big|\uI_{\alpha\beta}^\gamma(\vvro,|Z|)
\big|^2$. Without losing generality, we can assume that the observation point takes place on the $z$-axis so that $\varrho = \rho'$. Recalling $\uu\,^\gamma_{\alpha\beta}:\uu\,^\gamma_{\alpha\beta} = 1$, we have
\begin{equation}\label{uI2}
	\big|\uI\,^\gamma_{\alpha\beta} (\varrho, |Z|)\big|^2 \, \rightarrow\, \big|I\,^\gamma_\nu (\rho', |Z|)\big|^2 = (2\pi)^2 e^{-2\kappa''_a |Z|} \big[ I_0^\gamma + \nu I_1^\gamma + I_2^\gamma \big],
\end{equation}
where
\begin{align}\label{uI22}
	I_0^\gamma (\rho', |Z|)&= {|w_\gamma|^2 \over |Z|^2} e^{-w''_\gamma\rho'^2/|Z|}, \nonumber \\
	I_1^\gamma (\rho', |Z|) &= {1 \over \rho'^4} \left(1 + e^{-w''_\gamma\rho'^2/|Z|} \right) - {2 \over \rho'^4} e^{-w''_\gamma\rho'^2/(2|Z|)} \cos\left[w'_\gamma\rho'^2/(2|Z|)\right], \nonumber \\
	I_2^\gamma (\rho', |Z|) &= {2w''_\gamma \over \rho'^2|Z|} e^{-w''_\gamma\rho'^2/(2|Z|)} \left( e^{-w''_\gamma\rho'^2/(2|Z|)} - \cos\left[w'_\gamma\rho'^2/(2|Z|)\right] \right) \nonumber \\
	&- {2w'_\gamma \over \rho'^2|Z|} e^{-w''_\gamma\rho'^2/(2|Z|)} \sin[w'_\gamma\rho'^2/(2|Z|)].
\end{align}
In the above equations: $\kappa''_a$ is the imaginary part of $\kappa_a =\pm \sqrt{a^e_{t}a^m_{t}}$ with $\Im (\kappa_a)>0$; $w'_\gamma$ and $w''_\gamma$ are, in respective, the real and imaginary parts of $w_\gamma$ with $\gamma=p[s]$; $w_{p[s]}= \pm ({a^{e[m]}_z\over a^{e[m]}_t} \kappa_a)$ with $\Im(w_{p[s]})>0$; $\nu=2$ if $\alpha = m$ and $\beta=e$, otherwise $\nu=1$.
Now, we can include Eq.~(\ref{uI2}) within Eqs.~(\ref{fphrat}), (\ref{fprad}), and (\ref{fqrad}) to calculate heating rate, radiative heat power, and radiative heat flux, respectively. 

\section*{Appendix E: Calculating the uniaxial $p^{\alpha}_{rad}$}

Since the trace and integral operations for both electric and magnetic expressions in Eq.~(\ref{154-extra}) are the same, we can re-write them into one equation as follows 
\begin{align}\label{154}
	\Tr\left[ \ubG\,^{[\dagger]}_{ee[mm]}\cdot\uB_{s,e[m]}\right]_{_{\omega_0}}\!j_{_0}(\tau') &= \nonumber \\
	&\sum_{\gamma=p,s} \bigg[ \uG\,^{tt,\gamma\,[\dagger]}_{ee[mm]}(\vr,t,t'):\uB\,^{t}_{s,e[m]}(t') \nonumber \\	
	& + \uG\,^{zz,p[s\,\dagger]}_{ee[mm]}(\vr,t,t'):\uB\,^{z}_{s,e[m]}(t')\bigg]_{_{\omega_0}}\!j_{_0}(\tau')
\end{align}
Recalling Eq.~(\ref{EDGF}) and remembering that here $\vec{\varrho}=0$ and $|Z|=0$, we re-write the components of EDGF, i.e. $\uG\,^{ii,\gamma}_{\alpha\alpha}(\vr,t,t')$, with $i=t,z$, as follows 
\begin{align}\label{155}
	\uG\,^{ii,\gamma}_{\alpha\alpha}(t,t') &=  {\Dom \over (2\pi)^3}\int d\vk_{t} g^{\,ii,\gamma}_{\alpha\alpha}(\tau')\ue\,^{ii,\gamma}_{\alpha\alpha} \equiv \tilde{g}^{\,ii,\gamma}_{\alpha\alpha}(\tau')\ue\,^{ii,\gamma}_{\alpha\alpha},	\nonumber \\
	&g^{\,ii,\gamma}_{\alpha\alpha}(\tau')=C^{\,ii,\gamma}_{\alpha\alpha}j_{_0}(\tau')-{i\Dom \over 2} D^{\,ii,\gamma}_{\alpha\alpha}j_{_1}(\tau'),\qquad \tau'=\Dom(t-t')/2,
\end{align}
where $\ue\,^{ii,\gamma}_{\alpha\alpha}$, $C^{\,ii,\gamma}_{\alpha\alpha}$ and $D^{\,ii,\gamma}_{\alpha\alpha}$ are given by 
$\ubC$ and $\ubD$ \cite{60}. Recalling Eqs.~(\ref{141-1})-(\ref{141-2}), for the electric component, we have
\begin{align}\label{156}
	\left[\uG\,^{\gamma}_{ee}:\uB_{s}\!^{e}\right]&_{_{\omega_0}}j_{_0}(\tau') = \nonumber \\
	&\left[ \tilde{g}^{\,ii,\gamma}_{ee}\,\ue\,^{ii,\gamma}_{ee}:\overline{\overline{b}}\,^{e}_{0_i}j_{_0}(\tau') + \tilde{g}^{\,ii,\gamma}_{ee}\,\ue\,^{ii,\gamma}_{ee}:\overline{\overline{b}}\,^{e}_{1_i}(i\p_{t'}-i\p_{t'}^\dagger)\, j_{_0}(\tau')/2 \right]_{_{\omega_0}}, 
\end{align}
where $\overline{\overline{b}}\,^{e}_{n_i}=b\,^{e}_{n_i}\uI_{i}$ with $i=t,z$ and $n=0,1$. 
For each value of $\vk_{t}$ in Eq.~(\ref{155}), we have $\ue\,^{ii,\gamma}_{ee}:\overline{\overline{b}}\,^{e}_{n_i}=b^{\,e}_{n_i}$ and, thus, Eq.~(\ref{156}) reads
\begin{equation}\label{157}
	\left[\uG\,^{\gamma}_{ee}:\uB_{s}\!^{e}\right]_{_{\omega_0}}j_{_0}(\tau') = \left[ \tilde{g}^{\,ii,\gamma}_{ee}\,b^{\,e}_{0_i}j_{_0}(\tau') + \tilde{g}^{\,ii,\gamma}_{ee}\,b^{\,e}_{1_i}(i\p_{t'}-i\p_{t'}^\dagger)\, j_{_0}(\tau')/2 \right]_{_{\omega_0}}. 
\end{equation}
Remembering assumptions in Eqs.~(\ref{143})--(\ref{145}) and  the
integral relations between different orders of spherical Bessel
functions (cf. Eqs.~(\ref{151})-(\ref{152})) 
together with some manipulations, 
the electric (magnetic) component of uniaxial radiative power can be written as
\begin{align}\label{158-C}
	p^{e[m]}_{rad} &=+[-]{\Dom k_B T \over \pi^4} \Re\!\sum_{j,\gamma} \!\int\! k_{t}dk_{t} \left(1-\delta_{jz}\delta_{\gamma,s[p]}\right) \nonumber \\
	&\times \left[ \left( a^{e[m]}_{0_j}+ia^{e[m]}_{1_j}\dot{T}/2T\right) C^{jj,\gamma[\,*]}_{ee[mm]}+{\Dom^2 \over 12} a^{e[m]}_{1_j } D^{jj,\gamma[\,*]}_{ee[mm]} \right]_{_{\omega_0}} .
\end{align}

To integrate over $t'$ and $t''$ in the above equation and also in Eq.(\ref{phrat}), the following integral relations between different orders of spherical Bessel functions have been used:
\begin{equation}\label{151}
	\begin{aligned}
		I_1&=\int dt'dt''g^*(\tau_{a}'')g(\tau_{a}')j_{_0}(\tau''')={4\pi^2\over\Dom^2} 
		\left( \big|C^{\,a,\gamma}_{\alpha\beta}\big|^2 + {\Dom^2\over 12}\big|D^{\,a,\gamma}_{\alpha\beta}\big|^2\right), \\	
		I_2&=i\int dt'dt''g^*(\tau_{a}'')g(\tau_{a}')j_{_1}(\tau''')={4\pi^2\over 3\Dom}\Re\!\left(C^{\,a,\gamma}_{\alpha\beta} \, D^{\,a,\gamma\,*}_{\alpha\beta}\right), \\ 
		I_3&=i\int dt'dt''g^*(\tau_{a}'')h(\tau_{a}')j_{_0}(\tau''')=-{4\pi^2\over 3\Dom}\Re\!\left(C^{\,a,\gamma}_{\alpha\beta} \, D^{\,a,\gamma\,*}_{\alpha\beta}\right),	
	\end{aligned}
\end{equation}
where we employed the following orthogonality relations between different orders of spherical Bessel functions:
\begin{align}\label{152}
	&\int_{-\infty}^{\infty} j_{_0}(x)j_{_0}(x)=\pi, \qquad \qquad \qquad \quad \int_{-\infty}^{\infty} j_{_1}(x)j_{_1}(x)=\pi/3, \nonumber \\	
	&\int_{-\infty}^{\infty} j_{_0}(x)j_{_1}(x)/x=\pi/3, \qquad \qquad \quad \int_{-\infty}^{\infty} j_{_0}(x)j_{_0}(x\pm a)=\pi j_{_0}(a) \nonumber \\
	&\int_{-\infty}^{\infty} j_{_0}(x)j_{_1}(x\pm a)=\pm\pi j_{_1}(a), \qquad \int_{-\infty}^{\infty} j_{_1}(x)j_{_0}(x\pm a)=\mp\pi j_{_1}(a), \nonumber \\
	&\int_{-\infty}^{\infty} j_{_0}(x)j_{_2}(x)=0.
\end{align}

\section*{Appendix F: Calculation of the paraxial $q_{rad, z}$ }

To find the paraxial radiative flux in the direction $n \equiv z$ (cf. Eq.~(\ref{159})), first, we expand the expression $\uB_{s,\alpha}(t')$ in Eq.~(\ref{159-1}) and then we retain only the time-dependence of $G$'s. Thus, Eq.~(\ref{159-1}) takes on the following form:
\begin{align}\label{160}
	\ucG\,^{a}_{e\alpha}j_{_0}(\tau''')&=b^{\,\alpha}_{0_t}\uG\,^{a,\,\gamma}_{e\alpha}(t,t')\cdot\uG\,^{a,\,\gamma\dagger}_{m\alpha}(t,t'')j_{_0}(\tau''') \nonumber \\
	&+ b^{\,\alpha}_{1_t}/2\left[ \uG\,^{a,\,\gamma}_{e\alpha}(t,t')\cdot(i\p_{t'}-i\p_{t'}^\dagger)\uG\,^{a,\,\gamma\dagger}_{m\alpha}(t,t'')j_{_0}(\tau''')\right].
\end{align}

Now, remembering the assumptions stated in Sec. 4 (cf. Eqs.~(\ref{143})-(\ref{145})) and calculating the time derivative, Eq.~(\ref{160}) reads
\begin{align}\label{161}
	\ucG\,^{a}_{e\alpha}j_{_0}(\tau''') &=k_BT (a^\alpha_{0_t}+a^\alpha_{1_t}\dot{T}'/2T)\uG\,^{a,\,\gamma}_{e\alpha}(t,t')\cdot\uG\,^{a,\,\gamma\dagger}_{m\alpha}(t,t'')j_{_0}(\tau''') \nonumber \\
	&+ {i\Dom\over 4} a^\alpha_{1_t}k_BT \bigg[ \uH\,^{a,\,\gamma}_{e\alpha}(t,t')\cdot\uG\,^{a,\,\gamma\dagger}_{m\alpha}(t,t'')j_{_0}(\tau''') \nonumber \\
	&- \uG\,^{a,\,\gamma}_{e\alpha}(t,t')\cdot\uG\,^{a,\,\gamma\dagger}_{m\alpha}(t,t'')j_{_1}(\tau''') \bigg].
\end{align}	

Recalling Eqs.~(\ref{Comp-PEDGF1})-(\ref{Comp-PEDGF4}), (\ref{147}), and (\ref{149}), and integrating over $t'$ and $t''$ (cf. Eqs.~(\ref{151})-(\ref{152})), 
Eq.~(\ref{161}) changes to the following form 
\begin{align}\label{162}
	\int dt'dt''\ucG\,^{a}_{e\alpha}j_{_0}(\tau''') &= {k_B T \over (2\pi)^4} \uI\,^\gamma_{e\alpha}(\rho', |Z|) \cdot \uI\,^{\gamma\,\dagger}_{m\alpha}(\rho', |Z|)  \nonumber \\
	& \times \bigg[(a^\alpha_{0_t}+a^\alpha_{1_t}\dot{T}/2T) (C^{\,a,\gamma}_{e\alpha} C^{\,a,\gamma\,*}_{m\alpha} + {\Dom^2 \over 12} D^{a,\gamma}_{e\alpha} D^{a,\gamma\,*}_{m\alpha}  ) \nonumber \\
	&- {\Dom^2 \over 12} a^{\alpha}_{1_t} \left(C^{\,a,\gamma}_{e\alpha} D^{a,\gamma\,*}_{m\alpha}
	+ D^{a,\gamma}_{e\alpha} C^{\,a,\gamma\,*}_{m\alpha}\right) \bigg]_{_{\omega_0}}.
\end{align}
Including Eq.~(\ref{162}) in Eq.~(\ref{159}), $q_{rad}$ is reduced to
the following form
\begin{align}\label{164}
	q_{rad,n}(\vr,t;\omega_{0}) = {\Dom \over \pi^4}{k_B T \over (2\pi)^3} & \Re\!\sum_{\alpha,\gamma}
	\int d^3r'\left[\epsilon_{nkl} \hat{e}_k\cdot\uI\,^\gamma_{e\alpha}(\rho', |Z|) \cdot \uI\,^{\gamma\,\dagger}_{m\alpha}(\vR)\cdot\hat{e}_l\right] \nonumber \\
	&\times \bigg[ (a^\alpha_{0_t}+a^\alpha_{1_t}\dot{T}/2T) (C^{\,a,\gamma}_{e\alpha} C^{\,a,\gamma\,*}_{m\alpha} + {\Dom^2 \over 12} D^{a,\gamma}_{e\alpha} D^{a,\gamma\,*}_{m\alpha}  ) \nonumber \\
	&- {\Dom^2 \over 12} a^{\alpha}_{1_t} \left(C^{\,a,\gamma}_{e\alpha} D^{a,\gamma\,*}_{m\alpha}
	+ D^{a,\gamma}_{e\alpha} C^{\,a,\gamma\,*}_{m\alpha}\right) \bigg]_{_{\omega_0}}.
\end{align}
In order to calculate the expression in the first bracket in the above
equation, we set a system of orthogonal coordinates ($\hz,\htt,\hs$)
with unit vectors ($\hat{e}_n,\hat{e}_k,\hat{e}_l$) in Eq.~(\ref{159}). So, using Eqs.~(\ref{pcdhat}) and (\ref{intgl-1}), we obtain
\begin{align}\label{165}
	&\epsilon_{nkl}\hat{e}_k\cdot\uI\,^{s}_{e\alpha}(\vR)\cdot\uI\,^{s\dagger}_{m\alpha}(\vR)\cdot\hat{e}_l = 0, \nonumber \\ 
	&\epsilon_{nkl}\hat{e}_k\cdot\uI\,^{p}_{\,e\alpha}(\vR)\cdot\uI\,^{p\dagger}_{m\alpha}(\vR)\cdot\hat{e}_l = I^p_{\,e\alpha}(\vR)I^{\,p\,*}_{m\alpha}(\vR).
\end{align}
Now, by recalling Eq.~(\ref{intgl-2}), we have
\begin{equation}\label{intgl-3}
	I^p_{e\alpha}(\vR)I^{p\,*}_{m\alpha}(\vR)= \big|I^p_\nu(\rho', |Z|)\big|^2.
\end{equation}
Thus, the $z$-component of radiative heat flux reads
\begin{align}\label{166-D}	
	q_{rad,z}(\vr,t;\omega_{0}) = {\Dom \over \pi^4}{k_B T \over (2\pi)^3} &\Re\!\sum_{\alpha}
	\int d^3r'\,\big|I^p_\nu(\rho', |Z|)\big|^2 \nonumber \\
	&\times \bigg[ (a^\alpha_{0_t}+a^\alpha_{1_t}\dot{T}/2T) ( C^{a,p}_{e\alpha} C^{a,p\,*}_{m\alpha} + {\Dom^2 \over 12} D^{a,p}_{e\alpha} D^{a,p\,*}_{m\alpha} ) \nonumber \\
	& - {\Dom^2 \over 12} a^{\alpha}_{1_t} \left( C^{a,p}_{e\alpha} D^{a,p\,*}_{m\alpha}
	+ D^{a,p}_{e\alpha} C^{a,p\,*}_{m\alpha} \right) \bigg]_{_{\omega_0}}. 
\end{align}	

\section*{Appendix G: Thermal diffusion matrix equations}

According to the \textit{quadrature-difference method}, we rewrite the left-hand side of
Eq.~(\ref{HDE-2}) in terms of forward, central, and backward differences to calculate the second-order derivative of $T_n(z)$ with respect to $z$ (up to $O(h^4)$ with $h = (z_i - z_f)/N$ as the step size) as follows
\begin{align}\label{fcb-diff} 
	&\ \mbox{\textit{Forward difference}} \nonumber \\
	&T''_n(z_1) = {1 \over 12h^2}\left[\iota_1 T_n(z_1)+\iota_2 T_n(z_2) +\iota_3 T_n(z_3)+ \iota_4 T_n(z_4) + \iota_5T_n(z_5)\right], \nonumber \\
	&\ \mbox{\textit{Central difference}} \nonumber \\
	&T''_n(z_{m}) = {1 \over 12h^2}\left[\iota_6T_n(z_{m-2})+\iota_7T_n(z_{m-1})+\iota_8T_n(z_{m})+\iota_7T_n(z_{m+1})+\iota_6 T_n(z_{m+2})\right],  \nonumber \\
	& \qquad \qquad \qquad \qquad \qquad \qquad 1 < m < N-1, \nonumber \\
	&\ \mbox{\textit{Backward difference}} \nonumber \\
	&T''_n(z_{N-1}) = {1 \over 12h^2}\left[\iota_1T_n(z_{N-1})+\iota_2T_n(z_{N-2})+\iota_3T_n(z_{N-3})+\iota_4T_n(z_{N-4})+\iota_5T_n(z_{N-5})\right], \nonumber \\
\end{align}
where $\iota$'s coefficients are as follows
\begin{align}\label{iota-extra}
	&\iota_1=35, \qquad \iota_2 = -104, \qquad \iota_3=114, \qquad \iota_4=-56, \nonumber \\
	&\quad \iota_5=11, \qquad \iota_6=-1,\qquad \iota_7=16,\qquad \iota_8=-30.
\end{align}	
Regarding the Composite Simpson’s 1/3 rule, the right-hand side of Eq.~(\ref{HDE-2}) can be written as follows
\begin{align}\label{CSimp}
	\lambda_\rho \int_{z_i}^{z_f} M_n(z,z')&\,T_n(z')dz' = \lambda_\rho\,h/3 \times\big[M_n(z_m,z_i)\,T_n(z_i) + M_n(z_m,z_f)\,T_n(z_f) \nonumber \\
	&+ \sum_{l=1}^{N-1} \left[4M_n(z_m,z_{2l-1})T_n(z_{2l-1})+2M_n(z_m,z_{2l})T_n(z_{2l})\right]\big],
\end{align}
where $1 < m < N-1$ and $M_n$'s have closed-forms which we obtain below. 

To find a closed-form expression of $M_n (|Z|)$ with $|Z| = |z_m-z_l|$, let us take a closer look at Eqs.~(\ref{M_n}) and (\ref{uI2}). All terms there have Gaussian forms and we can easy compute them in terms of Gamma functions. Thus, we found three terms as follows:
\begin{align}\label{M_nn}
	M_n(|Z|) = (2\pi)^2 e^{-2\kappa^{''}_a |Z|} & \sum_{\alpha,\beta,\gamma, \omega_0}F^{a,c,st}_{\alpha\beta,\gamma}\int_{0}^{\infty} d\rho'\rho'^{n+1}e^{-\alpha_\sigma \rho'^{2}}  \nonumber \\
	&\times \left[ I_0^\gamma (\rho', |Z|)+\nu I_1^\gamma (\rho', |Z|)+I_2^\gamma (\rho', |Z|) \right].
\end{align}
Using the \textit{Mathematica} software to solve the above integral and after some algebra, we find
\begin{align}\label{M_nnn}
	M_n(|Z|) = 4\pi^2 &e^{-2\kappa^{''}_a |Z|} \nonumber \\
	& \times \sum_{\alpha,\beta,\gamma, \omega_0} F^{a,c,st}_{\alpha\beta,\gamma} \bigg[ \mathfrak{M}_{0,n}^\gamma (|Z|) + \nu\, \mathfrak{M}_{1,n}^\gamma (|Z|) + \mathfrak{M}_{2,n}^\gamma (|Z|) \bigg],
\end{align}
where 
\begin{align}\nonumber \label{M_n012-1-SM}
\mathfrak{M}_{0,n}^\gamma(|Z|) = {|w_\gamma|^2 \over |Z|^2} \, {\Gamma(n/2 + 1)  \over 2(\alpha_\sigma + w''_\gamma/|Z|)^{n/2+1} }, \quad \, &\nonumber \\
	\mathfrak{M}_{1,n = 0}^\gamma(|Z|) =  {1\over2} \,  \bigg[ \alpha_\sigma \mbox{log}(2\alpha_\sigma) - 2\,\Re\big(\alpha_\sigma - i w_\gamma&/2|Z|\big) \mbox{log}\big[2(\alpha_\sigma - i w_\gamma/2|Z|)\big] \nonumber\\ 
 	- (\alpha_\sigma +  w''_\gamma/|Z|) \mbox{log}\big[2(\alpha_\sigma + &w''_\gamma/|Z|)\big] \bigg], \nonumber \\
\mathfrak{M}_{1,n = 2}^\gamma(|Z|) = \,{1 \over 2} \mbox{log}\bigg[ 1 +  {(|w_\gamma|/2|Z|)^2 \over \alpha_\sigma(\alpha_\sigma + w''_\gamma/|Z|)} \bigg]&, \nonumber \\	
	\mathfrak{M}_{1,n}^\gamma(|Z|) = \, {1\over2} \, \Gamma(n/2 - 1) \bigg[  \left(\alpha_\sigma + w''_\gamma/|Z| \right)&^{1-n/2} \,-  2\alpha_\sigma\Re\big(\alpha_\sigma - iw_\gamma/2|Z|\big)^{-n/2} \nonumber \\
	  +\, \alpha_\sigma^{1-n/2} - \Re&\big[iw_\gamma/|Z| (\alpha_\sigma - iw_\gamma/2|Z|)^{-n/2}\,\big] \bigg] , \qquad n \neq 0, \, n \neq 2, \nonumber \\
	\mathfrak{M}_{2,n=0}^\gamma(|Z|) = -{1 \over |Z|} \bigg[ \Re[i w_\gamma \mbox{log}(\alpha_\sigma - i w_\gamma/2&|Z|)] \, + w''_\gamma \mbox{log} (\alpha_\sigma + w''_\gamma/|Z|)\bigg], \nonumber \\
	\mathfrak{M}_{2, n \neq 0}^\gamma(|Z|) = {\Gamma(n/2) \over |Z|}\bigg[ \Re\big[ iw_\gamma\big(\alpha_\sigma - i w_\gamma/&2|Z|\big)^{-n/2}\, \big] + w''_\gamma\big(\alpha_\sigma + w''_\gamma/|Z|\big)^{-n/2} \bigg].
\end{align}

Now, to find a matrix equation for computing each $T_n(z_m)$, without any
loss of generality, we consider the first five terms of the forward and backward differences in Eq.~(\ref{HDE-2}). Now, by denoting $T_n(z_m)$ as $T^n_m$ and $M_n(z_m,z_l)$ as $M^n_{ml}$ and recalling Eqs.(\ref{fcb-diff}) and (\ref{CSimp}), Eq.~(\ref{HDE-2}) reads: 
\begin{equation}\label{HDE-22}
	\begin{aligned}
		&\mbox{if}\; m = 1\; \mbox{or}\; N-1: \\
		&{1 \over 12h^2}\left[\iota_1T^n_m+\iota_2T^n_{m\pm1}+\iota_3T^n_{m\pm2}+\iota_4T^n_{m\pm3}+\iota_5T^n_{m\pm4}\right] + \\
		&(\lambda_Q - (n+1)\lambda_0) T^n_m + (n+2)^2 \lambda_z\,T^{n+2}_m + \lambda_\sigma\theta(n-2)\, T^{n-2}_m = \mathcal{M}^n_{m}, \\[2mm]
		&\mbox{and if}\;  2 \le m \le N-2:  \\ 
		&{1 \over 12h^2}\left[\iota_6T^n_{m-2} + \iota_7T^n_{m-1}+\iota_8T^n_{m} + \iota_6T^n_{m+1} +\iota_6 T^n_{m+2}\right] + \\
		&(\lambda_Q - (n+1)\lambda_0) T^n_m + (n+2)^2 \lambda_z T^{n+2}_m + \lambda_\sigma\theta(n-2)\, T^{n-2}_m = \mathcal{M}^n_{m}, \\
	\end{aligned}
\end{equation}
where
\begin{equation} 
	\mathcal{M}^n_{m} = {\lambda_\rho\,h \over 3} \left[ M^n_{m0} \,T^n_0 + M^n_{mN} \,T^n_N \,+ \sum_{l=1}^{N-1} (4M^n_{m\,2l-1} \,T^n_{2l-1} + 2M^n_{m\,2l} \,T^n_{2l} ) \right].
\end{equation}
Above, $\pm$ within terms with indices of $m\pm1$, etc, where $m=1$ or $(N-1)$, shows the forward or backward difference, respectively.
Now, by manipulating Eq.~(\ref{HDE-22}), one can write a matrix equation as follows:
\begin{align}\label{HDE-Mat-1}
	&\ubO\,^n \cdot \vbT^n + P^{n+2} \vbT^{n+2} + Q^{n-2} \vbT^{n-2} = \vbU^n, \qquad n = 0,1,2,...,& \nonumber \\
	&P^{n+2} = 12(n+2)^2\lambda_z h^2, \qquad Q^{n-2} = 12 \theta(n-2)\lambda_z h^2,
\end{align}
where $\vbT^n$ and $\vbU^n$ are $(N-1)$-dimensional column
vectors, whose elements are given by $T^n_1, T^n_2, ...,
T^n_{N-2}, T^n_{N-1}$ and $U^n_i = 4\lambda_\rho
[(M^n_{i0}+\delta_{i2})T^n_0 +
(M^n_{iN}+\delta_{i\,N-2})T^n_N]h^3$, respectively, with $\delta_{ij}$ being the Kronecker delta, $h$ being the step size, and $N$ being the total number of steps. 
The $(N-1) \times (N-1)$ matrix $\ubO\,^n$ has the following form
\begin{equation}\label{O-MatElm-E}
	\ubO\,^n =
	\begin{bmatrix}
		\vec{\bf{a}}^{\,n}_1 & \vec{\bf{b}}^{\,n}_2 & \vec{\bf{a}}^{\,n}_3 & \vec{\bf{b}}^{\,n}_4 & \vec{\bf{a}}^{\,n}_5 & \vec{\bf{b}}^{\,n}_6 & ... & \vec{\bf{b}}^{\,n}_{N-2} & \vec{\bf{a}}^{\,n}_{N-1} \\ 
	\end{bmatrix} + \uiO
\end{equation}
where $\vec{\bf{a}}^{\,n}_j$ and $\vec{\bf{b}}^{\,n}_j$ are the column vectors with the following elements:
\begin{align}\label{O-Elm-E}
	a^n_{ij} = -16\lambda_\rho M^n_{ij} h^3, \qquad	b^n_{ij} = -8\lambda_\rho M^n_{ij} h^3,  
\end{align} 
and $\uiO $, an $N-1$ $\times$ $N-1$ matrix, is given by 
\begin{equation}\label{O-MatElm_iota-E}
	\uiO =
	\begin{bmatrix}
		a_0 & \iota_2 & \iota_3 & \iota_4 & \iota_5 & 0 & ... & 0 & 0 \\ 
		\iota_7 & b_0 & \iota_7 & \iota_6 & 0 & 0 & ... & 0 & 0 \\
		\iota_6 & \iota_7 & b_0 & \iota_7 & \iota_6 & 0 & ... & 0 & 0 \\
		0 & \iota_6 & \iota_7 & b_0 & \iota_7 & \iota_6 & ... & 0 & 0 \\
		0 & 0 & \iota_6 & \iota_7 & b_0 & \iota_7 & ... & 0 & 0 \\
		. & . & . & . & . & . & ... & . & . \\
		. & . & . & . & . & . & ... & . & . \\
		. & . & . & . & . & . & ... & . & . \\
		0 & 0 &  ... & b_0 & \iota_7 & \iota_6 & ... & 0 & 0 \\
		0 & 0 &  ... & \iota_7 & b_0 & \iota_7 & ... & \iota_6 & 0 \\
		0 & 0 &  ... & \iota_6 & \iota_7 & b_0 & ... & \iota_7 & \iota_6 \\
		0 & 0 &  ... & 0 & \iota_6 & \iota_7 & ... & b_0 & \iota_7 \\
		0 & 0 &  ... & \iota_5 & \iota_4 & \iota_3 & ... & \iota_2 & a_0 \\
	\end{bmatrix},
\end{equation}
where $a_0 = \iota_1 + 12(\lambda_Q - (n+1)\lambda_0) h^2$ and $b_0 = \iota_8 + 12(\lambda_Q - (n+1)\lambda_0) h^2$. 

\section*{References}
\bibliographystyle{amsplain}

\begin{thebibliography}{99}
	\bibitem{JAMS-1} G. Kleinstein,  Quart. Appl. Math. {\bf 28}(4), pp. 527-537 (1971).
	\bibitem{1} H.B. Callen and T.A. Welton, Phys. Rev. {\bf 83}, 34 (1950). 
	\bibitem{2} D. Polder, and M. Van Hove, Phys. Rev. B {\bf 4}, 3303 (1971).

	\bibitem{3} S.M. Rytov, Y.A. Kravtsov, and V.I. Tatarskii, Principles of Statistical Radiophysics {\bf 249} (Springer, 1989). 
	\bibitem{4} J.B. Pendry, Radiative exchange of heat between nanostructures. J. Phys. Condens. Matter {\bf 11}, 6621 (1999). 
	\bibitem{5} J.-P. Mulet, K. Joulain, R. Carminati, and J.-J. Greffet, Appl. Phys. Lett. {\bf 78}, 2931 (2001). 
	\bibitem{6} A. Volokitin and B. Persson, Phys. Rev. B {\bf 63}, 205404 (2001). 
	\bibitem{7} A. Narayanaswamy, S. Shen, Sand G. Chen, Phys. Rev. B {\bf 78}, 115303 (2008).
	\bibitem{8} S. Shen, A. Narayanaswamy, G. Chen, Nano Lett. {\bf 9}, 2909 (2009).
	\bibitem{9} E. Rousseau, A. Siria, G. Jourdan, S. Volz, F. Comin, J. Chevrier, and J.-J. Greffet, Nat. Photonics {\bf 3}, 514 (2009).
	\bibitem{10} R.S. Ottens, V. Quetschke, S. Wise, A.A. Alemi, R. Lundock, G. Mueller, D.H. Reitze, D.B. Tanner, and B.F. Whiting, Phys. Rev. Lett. {\bf 107}, 014301 (2011).  
    \bibitem{PRB-2023} Y.-C. Hao, Y. Zhang, and H.-L. Yi, Phys. Rev. B {\bf 108}, 125431 (2023).
    \bibitem{Nature-2024} L. Tang, L.M. Corr\^{e}a, M. Francoeur, et al., Nature (2024).  https://doi.org/10.1038/s41586-024-07279-2.

	\bibitem{11} Z.M. Zhang, Nano/Microscale Heat Transfer (McGraw-Hill, New York, 2007).
	\bibitem{12} K. Park, and Z. Zhang, Front. Heat Mass Transfer {\bf 4}, 013001 (2013).

	\bibitem{JAMS-2} H. Ammari, B. Fitzpatrick, H. Lee, S. Yu, and H. Zhang, Quart. Appl. Math. {\bf 77}, 767 (2019).
	\bibitem{13} R. Venkatasubramanian, E. Siivola, T. Colpitts, and B. O’quinn, Nature {\bf 413}, 597 (2001). 
	\bibitem{14} P. Wang, et al., Nat. Mater. {\bf 2}, 402 (2003). 
	\bibitem{15} M. Laroche, R. Carminati, and J.-J. Greffet, J. Appl. Phys. {\bf 100}, 063704 (2006). 
	\bibitem{16} B. Tian, et al., Nature {\bf 449}, 885 (2007). 
	\bibitem{17} K. Park, S. Basu, W.P. King, and Z.M. Zhang, J. Quant. Spectrosc. Radiat. Transf. {\bf 109}, 305 (2008). 
	\bibitem{18} X. Liu, T. Tyler, T. Starr, A.F. Starr, N.M. Jokerst, and W.J. Padilla, Phys. Rev. Lett. {\bf 107}, 045901 (2011). 
	\bibitem{19} O. Ilic, M. Jablan, J.D. Joannopoulos, I. Celanovic, H. Buljan, and M. Solja\v{c}i\'{c}, Phys. Rev. B {\bf 85}, 155422 (2012). 
	\bibitem{20} S.I. Maslovski, C.R. Simovski, and S.A. Tretyakov, Phys. Rev. B {\bf 87}, 155124 (2013). 
	\bibitem{21} C. Simovski, S. Maslovski, I. Nefedov, and S. Tretyakov, Opt. Express {\bf 21}, 12, 14988 (2013). 
	\bibitem{22} P. Ben-Abdallah and S.-A. Biehs, Phys. Rev. Lett. {\bf 112}, 044301 (2014). 
	\bibitem{23} R. St-Gelais, B. Guha, L. Zhu, S. Fan, and M. Lipson, Nano Lett. {\bf 14}, 6971–6975 (2014). 
	\bibitem{24} C.R. Otey, L. Zhu, S. Sandhu, S. Fan, J. Quant. Spectrosc. Radiat. Transf. {\bf 132}, 3 (2014). 
	\bibitem{25} K. Kim, et al., Nature {\bf 528}, 387 (2015). 
	\bibitem{26} I. Latella, P. Ben-Abdallah, S.-A. Biehs, M. Antezza, and R. Messina, Phys. Rev. B {\bf 95}, 205404 (2017).
	\bibitem{27} L. Cui, et al., Nat. Commun. {\bf 8}, 14479 (2017). 
	\bibitem{28} R. Yu, A. Manjavacas, and F.J. Garc\'{i}a de Abajo, Nat. Commun. {\bf 8}, 2 (2017), 
	\bibitem{29} I. Latella, S.-A. Biehs, R. Messina, A.W. Rodriguez, and P. Ben-Abdallah, Phys. Rev. B {\bf 97} 035423 (2018).
	\bibitem{30} T. Inoue, K. Watanabe, T. Asano, and S. Noda, Opt. Express {\bf 26}, 2, 192 (2018).
	\bibitem{smartNFRHT} I. Latella, S.-A. Biehs, and P. Ben-Abdallah, Opt. Express {\bf 29} 24816 (2021).
	\bibitem{MMNat-2024} S. Safaei Jazi, I. Faniayeu, R. Cichelero, et al., Nat Commun 15, 1293 (2024). 
	\bibitem{MMApp-2023} A. O. Krushynska, S. Janbaz, J.H. Oh, M. Wegener, N.X. Fang, Appl. Phys. Lett. 123, 240401 (2023).
	
	\bibitem{31} K. Joulain, J.P. Mulet, F. Marquier, R. Carminati, and J.J. Greffet, Surf. Sci. Rep. {\bf 57}, 59 (2005).
	\bibitem{32} A.I. Volokitin and B.N.J. Persson, Rev. Mod. Phys.{\bf 79}, 1291 (2007).
	\bibitem{33} I.S. Nefedov and L.A. Melnikov, Appl. Phys. Lett. {\bf 105}, 161902 (2014).
	\bibitem{34} O.D. Miller, S.G. Johnson, and A.W. Rodriguez, Phys. Rev. Lett. {\bf 115} 204302 (2015).
	\bibitem{35} S.I. Maslovski, C.R. Simovski, S.A. Tretyakov, New J. Phys. {\bf 18}, 013034 (2016).
	\bibitem{36} S. Lang, G. Sharma, S. Molesky, P.U. Kr\"{a}nzien, T. Jalas, Z. Jacob, A.Yu. Petrov, and M. Eich, Sci. Rep. {\bf 7}, 13916 (2017).
	\bibitem{SupP-2022} R. Liu, C. Zhou, Y. Zhang, Z. Cui, X. Wu2, and H. Y, Int. J. Extrem. Manuf. {\bf 4}, 032002 (2022) .
	\bibitem{SupP-2024} S. Zhang, Y. Dang, X. Li, Y. Li, Y. Jin, P.K. Choudhury, J. Xu, Y. Ma, Nanoscale, {\bf 16}, 3, (2024).

	\bibitem{38} H. Mariji and S.I. Maslovski, in {\it Proceedings of SPIE Photonics Europe, 10671, Metamaterials XI}, edited by A.D. Boardman, A.V. Zayats, and K.F. MacDonald (SPIE, Strasbourg, 2018), p.~1067114. 

	\bibitem{39} R. Messina and P. Ben-Abdallah, Sci Rep. {\bf 3}, 1383 (2013).
	\bibitem{41} S. Molesky and Z. Jacob, Phys. Rev. B {\bf 91}, 205435 (2015).
	\bibitem{43} S. Basu, Y.B. Chen, and Z.M. Zhang, Int. J. Energy Research {\bf 31}, 689 (2007). 
	\bibitem{44} D.G. Cahill, P.V. Braun, G. Chen, D.R. Clarke, S. Fan, K.E. Goodson, P. Keblinski, W.P. King, G.D. Mahan, A. Majumdar, H.J. Maris, S.R. Phillpot, E. Pop, and L. Shi, Applied Physics Reviews {\bf 1}, 011305 (2014). 
	\bibitem{45} S. Basu, Z.M. Zhang, and C.J. Fu, Int. J. Energy Research {\bf 33} 1203 (2009).
	\bibitem{46} Y. Qi and M. C. McAlpine, Energy Environ. Sci. {\bf 3}, 1275 (2010).
	\bibitem{47} S. Narayana and Y. Sato, Phys. Rev. Lett. {\bf 108}(21), 214303 (2012) 
	\bibitem{48} G. Park, S. Kang, H. Lee, and W. Choi, Sci. Rep. {\bf 7}, 41000 (2017)

  \bibitem{SR-2022} T. Chen, J.H. Lin, Sci Rep {\bf 12}, 2734 (2022).
  \bibitem{MP-2024} D. Dubey, A.S. Mirhakimi, and M.A. Elbestawi, J. Manuf. Mater. Process. {\bf 8}, 40 (2024).	
	
     \bibitem{49} A. Lenert, D.M. Bierman, Y. Nam, W.R. Chan, I. Celanovic, M. Soljacic, and E.N. Wang, Nat. Nanotechnol. {\bf 9}, 126 (2014).
     \bibitem{50} B. Ko, D. Lee, T. Badloe, and J. Rho, Energies {\bf 12}, 89 (2019).


	\bibitem{51} C.C. Nadell, B. Huang, J.M. Malof, and W.J. Padilla, Deep learning for accelerated all-dielectric metasurface design, Opt. Express {\bf 27}(20), 27523 (2019).
	\bibitem{AI-MP} S. Lee, C. Park, J. Rho, Curr. Opin. Solid State Mater. Sci., {\bf 29} (2024).
	
	\bibitem{52} R. Schittny, M. Kadic, S. Guenneau, and M. Wegener, Phys. Rev. Lett. {\bf 110}(19), 195901 (2013). 
	\bibitem{53} N. Li, J. Ren, L. Wang, G. Zhang, P. H\''{a}nggi, and B. Li, Rev. Mod. Phys. {\bf 84}(3), 1045 (2012). 
	\bibitem{54} H. Han, L. G. Potyomina, A. A. Darinskii, S. Volz, and Y. A. Kosevich, Phys. Rev. B {\bf 89}(18), 180301 (2014). 
	\bibitem{55} J. Gomis-Bresco, et al, Nat. Commun. {\bf 5}, 4452 (2014).
	\bibitem{MicroTM} Z. Yan, M. Jin, Z. Li Z, G. Zhou, and L. Shui, Micromachines {\bf 10}(2), 89 (2019).
    \bibitem{MicroTM1-2024} P. Mittal, K.N. Nampoothiri, A. Jha, and S. Bansal, Int. J. Interact. Des. Manuf. (2024). 
    \bibitem{MicroTM2-2024} K.N. Nampoothiri, S. Bansal, A. Jha, P. Mittal, Eur. Phys. J. Spec. Top. (2024). 
    \bibitem{NanoTM} A. Moita, A. Moreira, J. Pereira, Symmetry {\bf 13}, 1362 (2021).
	\bibitem{56} J.-P. Mulet, K. Joulain, R. Carminati, and J.-J. Greffet, Microscale Thermophys. Eng. {\bf 6}, 209 (2002).
	\bibitem{57} J.B. Pendry, A.I. Fern\'{a}ndez-Dom\'{i}nguez, Y. Luo, and R. Zhao, Nat. Phys. {\bf 9}, 518 (2013).
	\bibitem{58} V. Chiloyan, J. Garg, K. Esfarjani, and G. Chen, Nat. Comm, {\bf 6}, 6755 (2015).
	\bibitem{59} G.W. Milton, M. Briane, and J. R. Willis, New J. Phys. {\bf 8}(10), 248 (2006).
	\bibitem{Ito-2017} K. Ito, K. Nishikawa, A. Miura, H. Toshiyoshi, and H. Iizuka, Nano Lett. {\bf 17}, 7 (2017).
	\bibitem{60} S.I. Maslovski and H. Mariji, Sci. Rep. {\bf 9}, 19980 (2019). 
	\bibitem{SPIE-Stanislav} S.I. Maslovski, in {\it Proceedings of SPIE Photonics Europe, 10671, Metamaterials XI}, edited by A.D. Boardman, A.V. Zayats, and K.F. MacDonald (SPIE, Strasbourg, 2018), p.~106710T-1. 
	\bibitem{Landau} L.D. Landau and E.M. Lifshitz, {\it Statical Physics}, translated by J.B. Sykes and M.J. Kearsley, Revised by L.P. Pitaevskii (London, Oxford, ELSEVIER SCIENCE and TECHNOLOGY, 1996).
	\bibitem{JAMS-3} Y. Bruned, F. Gabriel, M. Hairer, and L. Zambotti, J. Amer. Math. Soc. {\bf 35}, pp. 1-80 (2022).
	\bibitem{EMT} Choy, Tuck C., {\it Effective Medium Theory: Principles and Applications}, 2nd edn, International Series of Monographs on Physics (London, Oxford, 2015).
\end{thebibliography}

\end{document}